\documentclass[pra,preprint,showpacs]{revtex4}
\usepackage{times}
\usepackage{amsbsy,amssymb}
\usepackage{graphicx,color}

\def\one{{\mathchoice {\rm 1\mskip-4mu l} {\rm 1\mskip-4mu l} {\rm
1\mskip-4.5mu l} {\rm 1\mskip-5mu l}}}
\newcommand{\ket}[1]{|{#1}\rangle}
\newcommand{\bra}[1]{\langle{#1}|}

\newcommand{\ignore}[1]{}
\newcommand{\mComment}[1]{}
\newcommand{\gComment}[1]{}
\newcommand{\jComment}[1]{}
\newcommand{\rComment}[1]{}
\newcommand{\lComment}[1]{}
\renewcommand{\mComment}[1]{\textcolor{blue}{Manny: #1}}
\renewcommand{\gComment}[1]{\textcolor{red}{Gerardo: #1}}
\renewcommand{\jComment}[1]{\textcolor{green}{Jim: #1}}
\renewcommand{\rComment}[1]{\textcolor{magenta}{Ray: #1}}
\renewcommand{\lComment}[1]{\textcolor{purple}{Rolando: #1}}
\begin{document}

\title{Simulating Physical Phenomena by Quantum Networks}
\author{R. Somma}
\author{G. Ortiz}
\author{J. E. Gubernatis}
\author{E. Knill}
\author{R. Laflamme}
\affiliation{
Los Alamos National Laboratory, Los Alamos, NM 87545}
%\affiliation{
%Theoretical Division, 
%Los Alamos National Laboratory, Los Alamos, NM 87545}
%\author{E. Knill}
%\affiliation{ Computer and Computational Sciences Division, 
%Los Alamos National Laboratory, Los Alamos, NM 87545}

\date{\today}

\begin{abstract}
Physical systems, characterized by an ensemble of interacting 
elementary constituents, can be represented and studied by different
algebras of observables or operators. For example, a fully polarized 
electronic system can be investigated by means of the algebra 
generated by the usual fermionic creation and annihilation operators, 
or by using the algebra of Pauli (spin-$1/2$) operators. The 
correspondence between the two algebras is given by the Jordan-Wigner 
isomorphism. As we previously noted similar one-to-one mappings enable
one to represent any physical system in a quantum computer. 
In this paper we evolve and 
exploit this fundamental concept in quantum information processing 
to simulate generic physical phenomena by quantum networks. We give 
quantum circuits useful for the efficient evaluation of the physical
properties (e.g, spectrum of observables or relevant correlation
functions) of an arbitrary system with Hamiltonian $H$.
\end{abstract}

\pacs{Pacs Numbers: 3.67.Lx, 5.30.-d,} \columnseprule 0pt

\maketitle

\section{Introduction}
\label{section1}
 
\begin{figure}
\begin{center}
\includegraphics[angle=270,width=13.0cm,scale=1.0]{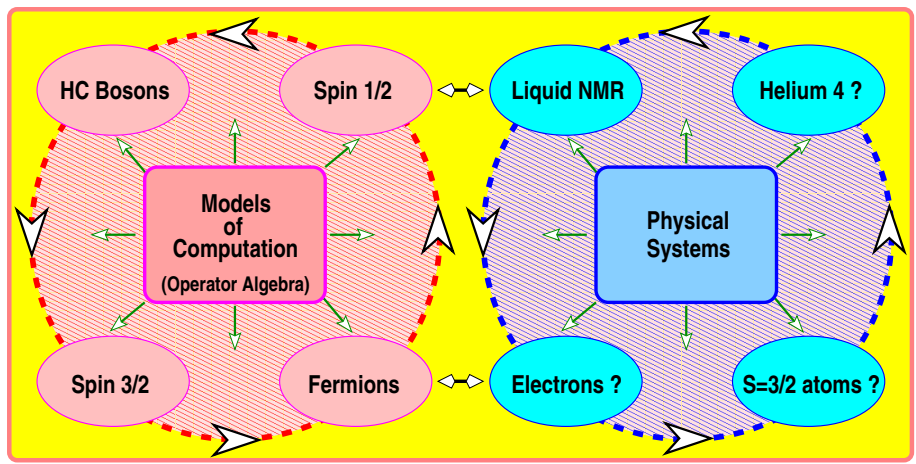}
\end{center}
\bigskip
\caption{Relationship between different models of computation
(with their associated operator algebras) and different physical systems.
Question marks refer to the present lack of a quantum computer 
device using the corresponding elementary physical components indicated 
in the box. Diamond-shaped arrows represent the natural connection 
between physical system and operator language, while arrows on the
circle indicate the existence of isomorphisms of $*$-algebras,
therefore, the corresponding simulation of one physical system by
another. A wave function view of this relationship is
given in~\cite{somaroo:qc1999a}.
}
\label{fig:1}
\end{figure}

A fundamental concept in quantum information processing is the
connection of a quantum computational model to a physical system by
transformations of closed operator algebras.  
%\mComment{Is it important to mention ``closed'' here?}  
The concept is a necessary one
because in quantum mechanics each physical system is naturally
associated with a language of operators (for example, quantum spin-1/2
operators) and thus to an algebra realizing this language (e.g., the
Pauli spin algebra generated by a family of commuting quantum spin-1/2
operators). Any quantum system defined by an algebra of operators
generated by a set of ``basic'' operators can be considered as a
possible model of quantum computation \cite{sfer}. The remarkable fact 
is that an
arbitrary physical system is simulatable by another physical system
(or quantum computer) whenever isomorphic mappings (embeddings)
between the two operator algebras exists. In each such case, an
important problem is to determine whether the simulation is efficient
(polynomial resource overhead) in terms of the ``basic'' generators.
For example, a nuclear spin (NMR) quantum computer is modeled as a
collection of quantum spin-1/2 objects and described by the Pauli
algebra.  It can simulate a system of $^4$He atoms (with space
discretized by a lattice) represented by the hard-core bosonic
algebra, and vice versa. In this case, the simulation is efficient.
Figure \ref{fig:1} summarizes this fundamental concept by giving a
variety of proposed physical models for quantum computers and
associated usable operator algebras. If one of these systems suffices
as the universal model of quantum computing, the mappings between the
operator algebras establish the equivalence of the other physical
models to it. This is one's intuitive expectation, and has a
well-established mathematical basis \cite{batista01}.

The mappings between algebras, between an algebra and a physical
system, and between physical systems are necessary in order to be able
to simulate physical systems using a quantum computer fabricated on
the basis of another system. However, this does not imply that the
simulation is efficiently implementable.  As we have previously
discussed \cite{sfer}, efficient quantum computation involves more
than having the ability to represent $2^N$ different items of classical
information so that the algebra of $N$ quantum bits (qubits) can be
isomorphically represented and quantum parallelism can be exploited.  It
is also insufficient for the mapping between operator algebras to be
easily and perhaps efficiently formalized symbolically.
%\mComment{What does ``map one to another with polynomial complexity'' mean?}  
For example, the physical system consisting of one boson in
$2^N$ modes is described using the language of ``transition''
operators that move the boson from one mode to the other.  Formally,
the Pauli matrices on $N$ qubits can be easily represented using the
transition operators, but the one-boson system is no more powerful
than classical wave mechanics.  This means that unless quantum
computers are not as powerful as is believed, there is no efficient
simulation of qubits by the one-boson system.

To be useful as a physics simulation device, a quantum computer must
answer questions about physical properties associated with real
physical systems. These questions are often concerned with the
expectation values of specific measurements of a quantum state evolved
from a specific initial state. Consequently, the initialization,
evolution, and measurement processes must all be implementable with
polynomial scaling \cite{sfer}. Often it is difficult to do. Further,
some classes of measurements, such as thermodynamic ones, still lack
well-defined workable algorithms \cite{terhal:qc1998a}.

On a classical computer, many quantum systems are simulated by the
Monte Carlo method \cite{qmcrev}. For fermions, the operation counts
of these Monte Carlo algorithms scale polynomially with the complexity
of the system as measured by the number of degrees of freedom, but the
statistical error scales exponentially (in time and in number of
degrees of freedom), making the simulation ineffective for large
systems. A quantum computer allows for the efficient simulation of
some systems that are impractical on a classical computer. In our
recent paper \cite{sfer} we discussed how to simulate a system of
spinless fermions by the standard model of a quantum computer, that
is, the model expressed in the language and algebra of quantum 
spin-1/2 objects (Pauli algebra). We also discussed how to make certain
physically interesting measurements. We demonstrated that the mapping
between algebras is a step of polynomial complexity and gave
procedures for initial state preparation, evolution, and certain
measurements that scaled polynomially with complexity.  The main focus
of the paper however was demonstrating that a particular problem for
simulating fermions on a classical computer, called the dynamical sign
problem, does not exist on a quantum computer. We are aware of at
least one case where the sign problem can be mapped onto an
NP-complete problem \cite{wiese}. This is the 3-SAT problem
\cite{farhi01}. Therefore, one cannot yet claim that a quantum
computer can solve ``all'' sign problems, otherwise one would claim
that one is solving all NP-complete problems and this has not been
rigorously established.

In this paper we continue to explore additional issues associated with
efficient and effective simulations of physical systems on a quantum
computer, issues which are independent of the particular experimental
realization of the quantum computer. We seek to construct quantum
network models of such computations. Such networks are sets of
elementary quantum gates to which we map our physical system. For
simplicity, we discuss these issues relative to simulating a system of
spin-1/2 fermions by the standard model of quantum computing. Our
discussion has obvious applications to the simulation of a system of
bosons (or any other particle statistics or, in mathematical terms,
any other operator algebra).  
%\mComment{Unnecessary explanation: Because of our previous work 
%\cite{sfer}, the restriction to fermions allows some reduction 
%in background presentation. } 
Specifically we
address issues discovered in our attempt to implement a (classical)
simulator of a network-based quantum computer and to conduct a quantum
computation on a physical system (NMR) with a small number of
qubits. On a classical computer the number of qubits simulatable is
limited by the exponential growth of the memory requirements.
Physically, we can only process information experimentally with
systems of a few qubits. Having the simulator permits a comparison
between theory and experiments likely to be realizable in the near
future. Overall, the main problems we address are how to reduce the
number of qubits and quantum logic gates needed for the simulation of
a particular physical phenomena, and how to increase the amount of
physical information measurable by designing efficient quantum
algorithms.

We organized the paper in the following manner: In
Section~\ref{section2} we summarize the quantum network representation
of the standard model of quantum computation, discussing both one- and
multi-qubit circuits. Then we summarize the connection between the
spin and fermion representations. In Section~\ref{section3}, we first
discuss the initialization, evolution, and measurement processes. In
each case we define procedures simpler than the ones presented in our
previous paper, greatly improving the efficiency with which they can
be done.  Greatly expanded are the types of measurements now
possible. For example, besides certain correlation functions, the
spectrum of operators, including the energy operator, can now be
obtained.  Our application of this technology to a system of fermions
on a lattice and the construction of a simulator is discussed in
Section~\ref{section4}. The Hubbard model is used as an example. We
conclude with a summary and a discussion of areas needing additional
work. The appendices contain technical points about the preparation of
coherent and correlated states and the use of the discrete classical
Fourier transformation.

\section{Quantum Network Representation of Physical Phenomena}
\label{section2}

It is the formal connection between models of computation and physical
systems described in the Introduction that allows one to simulate
quantum phenomena with a quantum computer. Simulation is realized
through a quantum algorithm that consists of unitary operations and
measurements. One of the objectives is to accomplish simulation
efficiently, i.e, with polynomial complexity. The hope is that quantum
simulation is ``more'' efficient (less resources) than classical
simulation and there are examples that support such hope
\cite{sfer}. In the following subsections we summarize the main
concepts in the
representation of physical phenomena by quantum networks.

\subsection{Standard Model}

In the standard model of quantum computation, the quantum bit, or {\em
qubit}, is the fundamental unit. A qubit's state $\ket{\sf a}$ is a
linear combination of the states $\ket{0}$ and $\ket{1}$ (e.g, a spin
1/2 with $\ket{0}=\ket{\uparrow}, \ \ket{1}=\ket{\downarrow}$):
\begin{equation}
\ket{\sf a}=a \ \ket{0}+ b \ \ket{1} \ ,
\end{equation}
where the complex numbers $a$ and $b$ are normalized to unity:
$|a|^2+|b|^2=1$.

\begin{figure}
\begin{center}
\includegraphics[angle=0,width=13.0cm,scale=1.0]{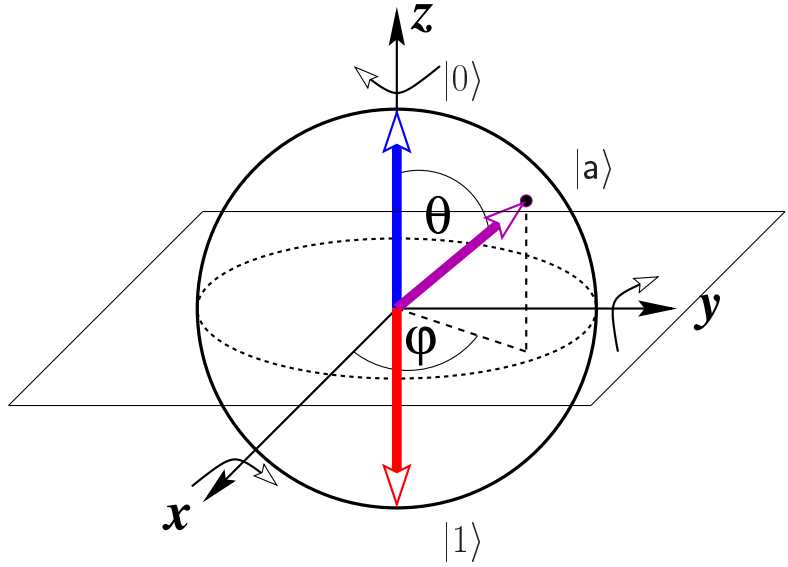}
\end{center}
\bigskip
\caption{Bloch-Sphere representation of a one qubit state parametrized as
$\ket{\sf a}=\cos\frac{\theta}{2} \ket{0} +e^{i\varphi}\sin\frac{\theta}{2}
\ket{1}$. The curved
arrows indicate the sign of rotation of $e^{i \frac{t}{2}\sigma_\mu}
=R_\mu(-t)$ about the particular axis $\mu$. Our (arrow) color convention is: 
$\ket{0} \rightarrow$ blue; $\ket{1} \rightarrow$ red; other linear
combinations $\rightarrow$ magenta.}
\label{fig:2}
\end{figure}

Assigned to each qubit are the identity matrix $\one$ and the
Pauli matrices $\sigma_x$, $\sigma_y$ and $\sigma_z$:
\begin{equation}
\one = \pmatrix{1&0 \cr 0&1 \cr} \ , \ \sigma_x= \pmatrix{ 0&1 \cr
1&0 \cr} \ , \ \sigma_y=\pmatrix{ 0&-i \cr i&0 \cr} \ , \
\sigma_z=\pmatrix{ 1&0 \cr 0&-1 \cr} \ , \label{eq:pauli}
\end{equation}
or equivalently $\one$, $\sigma_\pm=\frac{1}{2} (\sigma_x {\pm} i
\sigma_y)$, and $\sigma_z$. In this particular representation, the
states $\ket{0}$ and $\ket{1}$ are the vectors:
\begin{equation}
\ket{0}=\pmatrix{1\cr 0 \cr} \mbox{ and } \ket{1}=\pmatrix{0\cr 1
\cr} \ ,
\end{equation}
and the Bloch-Sphere (Fig. \ref {fig:2}) provides a convenient
three-dimensional real space representation of the single qubit
state $\ket{\sf a}$, which can be parametrized as
$\ket{\sf a}=\cos \frac{\theta}{2}\ket{0}+e^{i\varphi}\sin 
\frac{\theta}{2}\ket{1}$.

For a system of $n$ qubits, the mathematical representation of the
standard model is defined by a closed $*$-algebra (Pauli algebra) generated
by the operators $\sigma_{\mu}^{j}$ ($\mu= x$, $y$, or $z$) that act on
the $j^{th}$ qubit:
\[
\sigma^j_\mu = \overbrace{\one \otimes \one \otimes \cdots \otimes
\underbrace{\sigma_\mu}_{j^{th}\ \mbox{factor}} \otimes \cdots
\otimes \one}^{n\ \mbox{factors}} \ ,
\]
where $\otimes$ represents a Kronecker product. From these 
definitions, the resulting commutation relations are
\begin{eqnarray}
{[}\sigma_{\mu}^{j},\sigma_{\nu}^{j}{]}_{+}&=& 2\delta_{\mu\nu} \\
{[}\sigma_{\mu}^{j},\sigma_{\nu}^{k}{]}_{-}&=&
2i\delta_{jk}\epsilon_{\mu\nu \lambda}\sigma^{j}_{\lambda} \ ,
\end{eqnarray}
where $[A,B]_{\pm}=AB{\pm}BA$, and $\epsilon_{\mu\nu\lambda}$ is the
totally anti-symmetric Levi-Civita symbol. The time evolution of an $n$
qubit system is described by the unitary operator $\hat{U}(t)=e^{-iHt}$,
where $H$ represents the time-independent Hamiltonian of the system. In
turn, $\hat{U}(t)$ is easily expressible in terms of the Pauli matrices
$\sigma_{\mu}^j$ since they and their products form an
operator basis of the algebra. 

The most general unitary operator ${U}$ on a single qubit can be written
as
\begin{equation}
{U}=e^{i\alpha}R_{z}(\beta)R_{y}(\gamma)R_{z}(\delta) \ ,
\end{equation}
where $\alpha$, $\beta$, $\gamma$, and $\delta$ are real numbers, and 
$R_{\mu}(\vartheta)=e^{-i\frac{\vartheta}{2}\sigma_{\mu}}$ are
rotations in spin space by an angle $\vartheta$ about the $\mu$ axis. 
Although this decomposition is not unique, it is important because any
one qubit evolution is seen to be a combination of simple rotations (up
to a phase) about the $\mu= x$, $y$ or $z$ axis.

%In the standard model, the unitary operator $\hat{U}(t)=e^{-iHt}$ describing
%the temporal evolution of the $n$ qubits is easily expressible in terms
%of the Pauli matrices. A most general unitary operator ${U}$ on a single
%qubit can be written 
%\begin{equation}
%{U}=e^{i\alpha}R_{z}(\beta)R_{y}(\gamma)R_{z}(\delta) \ ,
%\end{equation}
%where $\alpha$, $\beta$, $\gamma$, and $\delta$ are real numbers, and 
%$R_{\mu}(\vartheta)=e^{-i\frac{\vartheta}{2}\sigma_{\mu}}$ are
%rotations in spin space about the $\mu$ axis. Although this
%decomposition is not unique, it is important because any one qubit
%evolution is seen as a combination of simple rotations (up to a phase)
%about the $x$, $y$ or $z$ axis.

In multi-qubit operations, any unitary operation ${U}$ can be
decomposed (up to a phase) as ${U}=\prod\limits_{l}U_{l}$, where $U_{l}$
are either single qubit rotations $R_{\mu}(\vartheta)$ in the $n$-qubit
space or two qubit interactions $R_{z^{j},z^{k}}(\omega) =
e^{i{\omega}\sigma_{z}^{j} \sigma_{z}^{k}}$ in the same space ($\omega$
is a real number) \cite{barenco:qc1995a,divincenzo:qc1995a}. These one
qubit rotations and two qubit interactions constitute the elementary
gates of the quantum computer in the network model.
%At a
%given time $t$, these one qubit rotations and two qubit interactions
%constitute the elementary gates of the quantum computer in the network
%model. Such a computer is a network temporally sequencing of these
%gates, based on the evolution being factorizable as $U(t)\equiv
%U(t,0)=U(t,t-\Delta t)U(t-\Delta t,t-2\Delta t)\cdots U(\Delta t,0)$.

%The basic relation is
%\begin{equation}
%e^{-i{\theta}\Sigma}={[}\cos \theta-i\sin \theta \Sigma{]}
%\end{equation}
%where $\theta$ is a real number and $\Sigma=\prod_{j,\mu}\sigma_{\mu}^{j}$. 

\subsection{Quantum Network}
\label{section2b}
We now describe some common one and two qubit gates, some quantum
circuits, and one pictorial way to represent them. The motivation for
this elementary subsection is to prepare the grounds for the quantum 
network simulation of a physical system developed in
Section~\ref{section3} which is more technically involved.

The goal is to represent any unitary operation (evolution) as a product
of one and two qubit operations. Although here we use the algebra of 
the Pauli matrices  (standard model), for a different model of
computation we should change the set of elementary gates, but the
general methodology remains the same. For instance, if the evolution
$\hat{U}(t)=e^{-iHt}$ is due to the Hamiltonian
\begin{equation}
H=H_{x}+H_y=\bar{\alpha} \ \sigma_x^1\sigma_z^2 \cdots\sigma_z^{j-1}
\sigma^j_x+\bar{\beta} \ \sigma_y^1\sigma_z^2 \cdots\sigma_z^{j-1}
\sigma^j_y \ ,
\end{equation}
where $\bar{\alpha}$ and $\bar{\beta}$ are real numbers, we write
$\hat{U}(t)$ as  $e^{-iH_{x}t} e^{-iH_{y}t}$ because $[H_{x},H_{y}]_{-} = 0$.
To decompose this into one and two qubit operations, we take the
following steps: We first note that the unitary operator
\begin{equation}
U_{1}=e^{i\frac{\pi}{4}\sigma_{y}^1}=\frac{1}{\sqrt{2}}
\left[{\one} +i \sigma_y^1\right]
\end{equation}
takes $\sigma_{z}^{1} \rightarrow \sigma^1_x$, i.e., $U_1^\dagger
\sigma_z^1 U_1 = \sigma_x^1$, so $U_1^{\dagger} e^{i{\bar{\alpha}}
\sigma_z^1} U_1 = e^{i{\bar{\alpha}}\sigma_x^1}$. Next we note that the
operator
\[
U_2=e^{i{\pi\over 4}\sigma_z^1\sigma_z^2}= \frac{1}{\sqrt{2}}
\left [{\one}+ i \sigma_z^1 \sigma_z^2 \right]
\]
takes $\sigma^1_x \rightarrow \sigma^1_y\sigma^2_z$, so $U_2^\dagger
e^{i{\bar{\alpha}} \sigma_x^1} U_{2} = e^{i{\bar{\alpha}} \sigma_y^{1}
\sigma_z^2}$. Then we note that
\[
U_3=e^{i{\pi\over 4}\sigma_z^1\sigma_z^3}
\]
takes $\sigma^1_y\sigma^2_z \rightarrow -\sigma^1_x \sigma^2_z
\sigma^3_z$. By successively similar steps we easily build the required
string of operators: $\sigma_x^1\sigma_z^2 \cdots\sigma_z^{j-1}
\sigma^j_x$ and also $e^{i{\bar{\alpha}}\sigma_x^1\sigma_z^2
\cdots\sigma_z^{j-1} \sigma^j_x}$ (up to a global phase):
\begin{equation}
U_{k}^{\dagger}\cdots U_{2}^{\dagger}U_{1}^{\dagger}
e^{i{\bar{\alpha}}\sigma_z^1} U_{1} U_{2} \cdots U_{k} =
e^{i{\bar{\alpha}}\sigma_x^1\sigma_z^2 \cdots\sigma_z^{j-1} \sigma^j_x}
\ 
\end{equation}
where the integer $k$ scales polynomially with $j$ (in this particular
case the scaling is linear). In a similar way, we decompose the
evolution $e^{-iH_{y}t}$. Multiplying both decompositions, we have the
total decomposition of the evolution operator $\hat{U}(t)$.
See~\cite{price:qc1999a,somaroo:qc1998a} for complete treatments
of these techniques.

\subsubsection{Single Qubit Circuits}
\begin{figure}
\begin{center}
\includegraphics[angle=0,width=13.0cm,scale=1.0]{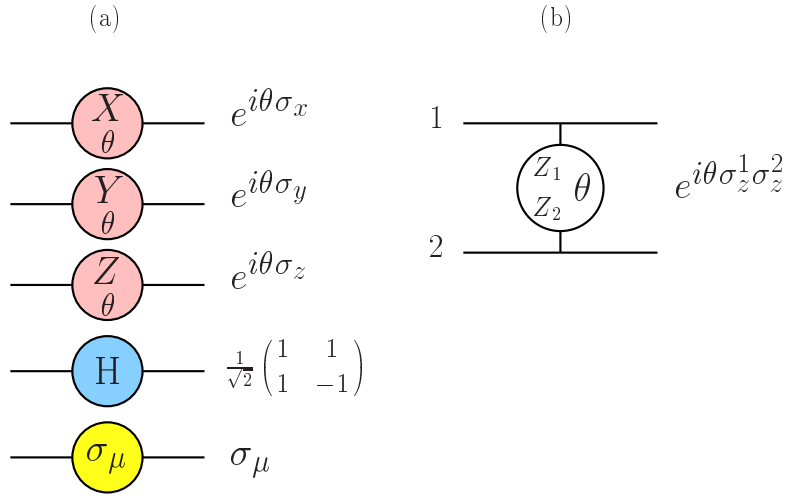}
\end{center}
\bigskip
\caption{(a) Some one qubit elementary gates and (b) a two qubit elementary gate.}
\label{fig:3}
\end{figure}

\begin{figure}
\begin{center}
\includegraphics[angle=0,width=13.0cm,scale=1.0]{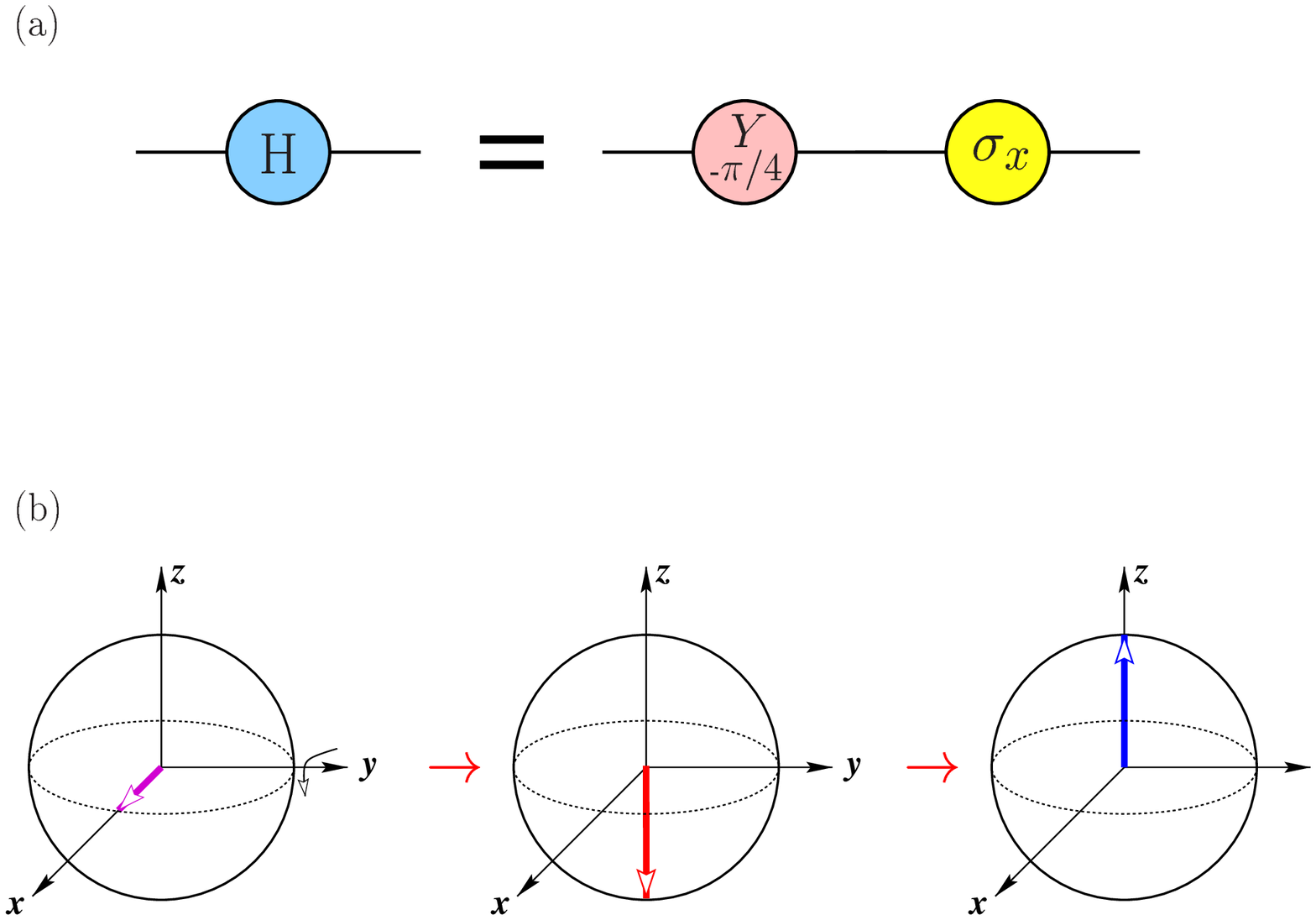}
\end{center}
\bigskip
\caption{(a) Hadamard gate decomposition and (b) Bloch-Sphere
representation of a Hadamard gate applied
to the state $\ket{+}$.}
\label{fig:4}
\end{figure}

In Fig.~\ref{fig:3}a we show examples of several elementary one qubit
gates. (Notice that $e^{i \theta \sigma_\mu}=R_{\mu}(-2\theta)$.) 
%\mComment{Unfortunately, the sign convention is opposite to
%the one I use, i.e. the $\theta$ rotation around $y$ is
%$e^{-i\sigma_y\theta/2}$. It may not be worth fixing this, but we
%should note the fact explicitly.}
Each gate applies one or more unitary operations
$R_{\mu}(\vartheta)$ to the qubit (the $\sigma_{\mu}$ gates apply a
$R_{\mu}(\pi)$ rotation up to a phase: $\sigma_{\mu} =
ie^{-i\frac{\pi}{2}\sigma_{\mu}}$). Also, in Fig.~\ref{fig:3}a we show
the Hadamard gate H. The action of this gate on the state of one qubit
$\ket{\sf a}$ is:
\[ 
\mbox{H} \left\{ \begin{array}{c}
 \ket{0} \leftrightarrow \ket{+}=(\ket{0}+\ket{1})/\sqrt{2} \\
 \ket{1} \leftrightarrow \ket{-}=(\ket{0}-\ket{1})/\sqrt{2}
\end{array}\right. \ .
\]
In this way, the Hadamard gate admits the matrix representation:
\begin{equation}
\mbox{H}=\frac{1} {\sqrt{2}}\pmatrix{1 & 1 \cr 1 & -1} \ .
\end{equation}
%From (\ref{eq:paulii}) and the identity $\sigma_{x}=i\sigma_{z}
%\sigma_{y}$, we express $\mbox H$
In terms of the Pauli matrices 
\begin{equation}
\mbox{H} = \frac{1}{\sqrt{2}}[\sigma_{x}+\sigma_{z}] =
ie^{-i \frac{\pi}{2} \sigma_{x}}e^{-i \frac{\pi}{4} \sigma_{y}} \ .
\end{equation}
In Fig. \ref{fig:4}a we show the decomposition of the {\rm H} gate into
single qubit rotations, and its application to the Bloch-Sphere
representation of the state $\ket{+}$ is shown in Fig.~\ref{fig:4}b. 
The convention for quantum circuits is each horizontal line represents
the time evolution of a single qubit and the time axis of the evolution
increases from left to right.

\subsubsection{Multiple Qubit Circuits}
\begin{figure}
\begin{center}
\includegraphics[angle=0,width=13.0cm,scale=1.0]{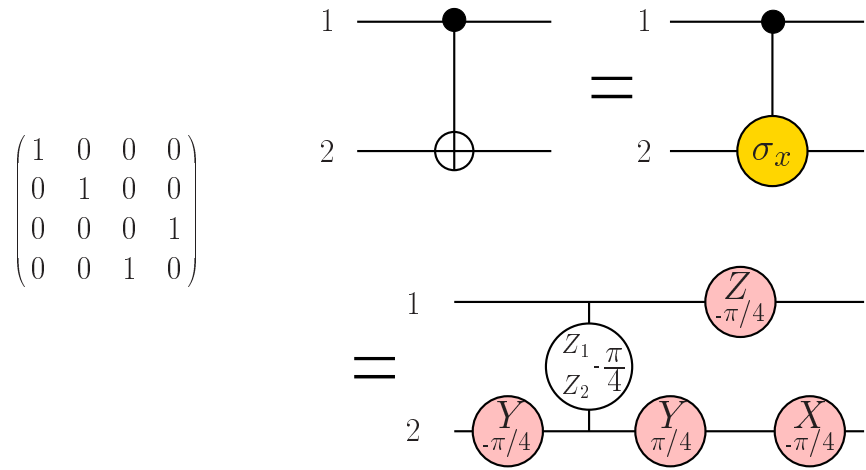}
\end{center}
\bigskip
\caption{C-NOT gate decomposition and its matrix representation. The
control qubit is 1. Note that the last circuit realizes the C-NOT matrix
operation up to a global phase $e^{-i \frac{\pi}{4}}$.}
\label{fig:5}
\end{figure}

\begin{figure}
\begin{center}
\includegraphics[angle=0,width=13.0cm,scale=1.0]{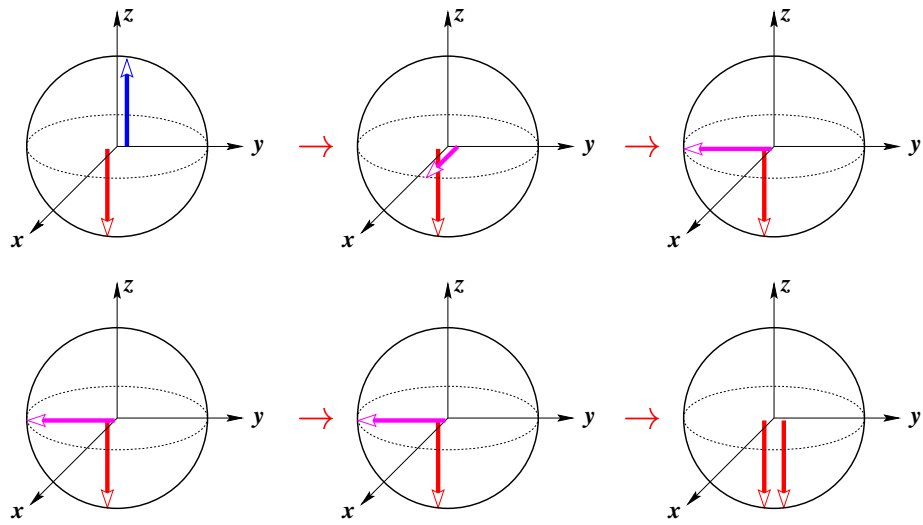}
\end{center}
\bigskip
\caption{Bloch-Sphere representation of the state obtained by the
C-NOT gate applied to the ``classical'' state $\ket{10}$.
The sequence of elementary operations is the same as fig. \ref{fig:5}
(time flows from left to right with the lower row continuing the 
upper one). For each Bloch-Sphere the two arrows indicate the states 
of the two qubits, with the left representing qubit one.}
\label{fig:6}
\end{figure}

We now give examples of multi-qubit operations. Again the goal is to
represent them as a combination (up to a phase) of single qubit
rotations $R_{\mu}(\vartheta)$ and two qubit interactions
$R_{z^{j},z^{k}}(\omega)= e^{i{\omega}\sigma_{z}^{j}\sigma_{z}^{k}}$
(the gate  for the $R_{z^{j},z^{k}}(\omega)$ is shown in
Fig.~\ref{fig:3}b). To illustrate this, we consider the circuit shown
in Fig. \ref{fig:5}. This is a two qubit controlled-NOT (C-NOT) gate
which acts as follows:
\[
\mbox{C-NOT}
\left\{\begin{array}{c}
\ket{00}\rightarrow\ket{00} \\
\ket{01}\rightarrow\ket{01} \\
\ket{10}\rightarrow\ket{11} \\
\ket{11}\rightarrow\ket{10}
\end{array}\right. \ .
\]
Here, the first qubit is the control qubit (the controlled operation on
its state $\ket{1}$ is represented by a solid circle in Fig.
\ref{fig:5}). We see that if the state of the first qubit  is $\ket{0}$
nothing happens, but if the first qubit is in $\ket{1}$, then the state
of the second qubit is flipped. Because  $\sigma_x^2$ is the unitary
operator that flips the second qubit (see Fig.  \ref{fig:5}), the
decomposition of the C-NOT operation into one and two qubits
interaction is 
\begin{equation}
\label {cnotdec}
 \mbox{C-NOT: }
e^{i \frac{\pi}{4}} e^{-i \frac{\pi}{4} \sigma_{z}^{1}}
 e^{-i \frac{\pi}{4} \sigma_{x}^{2}}
 e^{ i \frac{\pi}{4}\sigma_{z}^{1}\sigma_{x}^{2}} =
e^{i \frac{\pi}{4}} e^{-i \frac{\pi}{4} \sigma_{z}^{1}}
 e^{-i \frac{\pi}{4}\sigma_{x}^{2}}
 e^{ i \frac{\pi}{4}\sigma_{y}^{2}}
 e^{-i \frac{\pi}{4}\sigma_{z}^{1}\sigma_{z}^{2}}
 e^{-i \frac{\pi}{4}\sigma_{y}^{2}} \ .
\end{equation}
From Eq. \ref{cnotdec} we can see that a single controlled operation
becomes a greater number (in this case 4) of one and two qubits
operations. In Fig. \ref{fig:5} we also show the circuit representing
this decomposition, while in Fig. \ref{fig:6} we show the C-NOT gate
applied to the state $\ket{10}$ in the Bloch-Sphere representation.
Because of the control qubit being in the state $\ket{1}$, the second
qubit is flipped.

\begin{figure}
\begin{center}
\includegraphics[angle=0,width=13.0cm,scale=1.0]{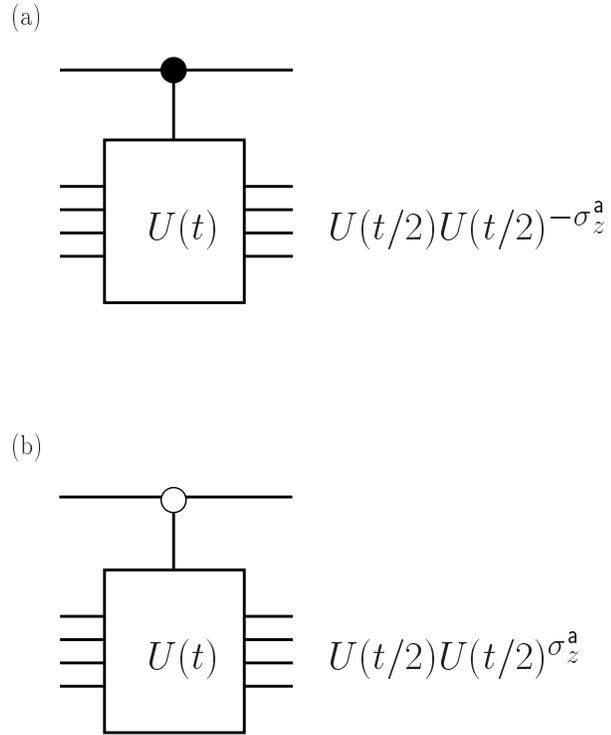}
\end{center}
\bigskip
\caption{(a) C-$U$ operation with the state of the control qubit ${\sf
a}$ being in
$\ket{1}_{\sf a}$ and (b) C-$U'$ operation controlled 
with the state $\ket{0}_{\sf a}$. (See text for notation.)}
\label{fig:7}
\end{figure}

A generalization of the C-NOT gate is the controlled-$U$ (C-$U$) gate,
where $U$ is a unitary operator acting on a multi-qubit state
$\ket{\Psi_{s}}$:
\[ 
\mbox{C-$U$} \left\{\begin{array}{l}
\ket{0}_{\sf a}\otimes\ket{\Psi_{s}}\rightarrow\ket{0}_{\sf a}\otimes\ket{\Psi_{s}} \\
\ket{1}_{\sf a}\otimes\ket{\Psi_{s}}\rightarrow\ket{1}_{\sf a}\otimes
{\biggl[}U\ket{\Psi_{s}}{\biggr]}
\end{array}\right. \ .
\] 
Mathematically, for $U(t)=e^{-i\hat{Q}t}$ ($\hat{Q}$ is Hermitian), 
the operational representation of the C-$U$ gate is: 
$U(t/2)U(t/2)^{-\sigma_{z}^{\sf a}}$ ($U(t)^{-\sigma_{z}^{\sf a}}=
e^{i\hat{Q}\otimes{\sigma_{z}^{\sf a}t}}$), where
${\sf a}$ is the control qubit (Fig.  \ref{fig:7}a). 
Similarly, one can use $\ket{0}_{\sf a}$ as the
control state to define the C-$U^\prime$ gate illustrated in
Fig.~\ref{fig:7}b.  In order to describe the C-$U$ and C-$U^\prime$
gates as a combination of single qubit rotations and two qubits
interactions, we have to decompose the operators $U(t/2)$ and
$U(t/2)^{\sigma_z^{\sf a}}$ into such operations. C-$U$ can then be
expressed as a sequence of conditional one and two qubit
rotations. The latter can be further decomposed into one and two qubit
rotations using the techniques of~\cite{barenco:qc1995a}.

\subsection{Spin-Fermion connection}
\label{section2c}
To simulate fermionic systems with a quantum computer that uses the
Pauli algebra, we first map the fermionic system into the standard
model \cite{sfer,bravyi:qc2000a}. The commutation relations for (spinless) fermionic
operators $a_{j}$ and $a_{j}^{\dagger}$ (the destruction and creation
operators for mode $j$) are
\begin{eqnarray}
{[}a_{j}^{\dagger},a_{k}{]}_{+}&=&\delta_{jk} \ , \\
{[}a_{j}^{\dagger},a_{k}^{\dagger}{]}_{+}&=&0 \ .
\end{eqnarray}
We map this set of operators to one expressed in terms of the
$\sigma_{\mu}^j$'s in the following way:
\begin{eqnarray}
a_j^{\;} &\rightarrow& \left( \prod_{l=1}^{j-1} -\sigma^l_z
\right) \sigma^j_- = (-1)^{j-1} \ \sigma^1_z \sigma^2_z \cdots
\sigma^{j-1}_z \sigma^j_-
\ , \nonumber \\
a^{\dagger}_j &\rightarrow& \left( \prod_{l=1}^{j-1} -\sigma^l_z
\right) \sigma^j_+ = (-1)^{j-1} \ \sigma^1_z \sigma^2_z \cdots
\sigma^{j-1}_z \sigma^j_+  \ . \nonumber
\end{eqnarray}
Obviously, for the fermionic commutation relations to remain satisfied,
the operators $\sigma_{\mu}^j$ must satisfy the commutation relations
of the Pauli matrices, so a representation for the operators
$\sigma_{\mu}^j$ are the Pauli matrices.

\begin{figure}
\begin{center}
\includegraphics[angle=0,width=13.0cm,scale=1.0]{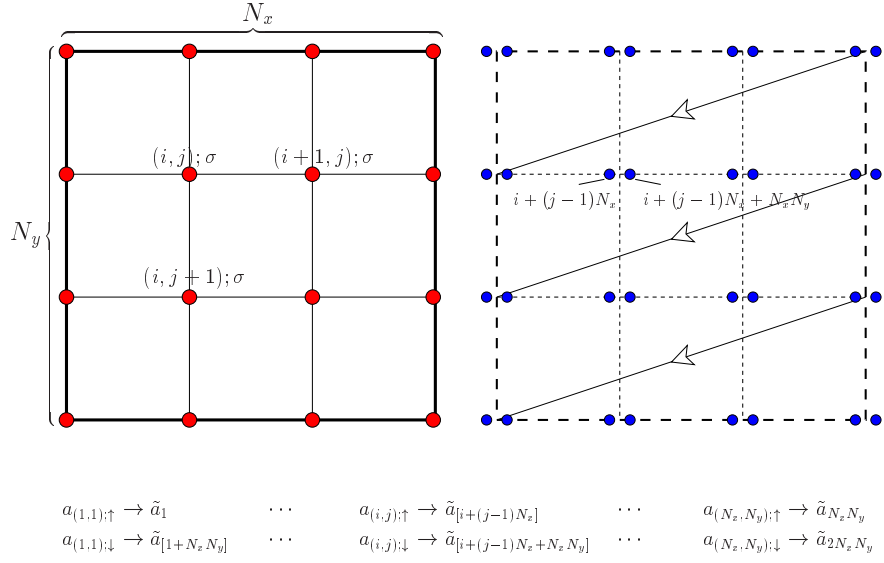}
\end{center}
\bigskip
\caption{Mapping used to connect the labels of a two-dimensional 
$N_x \times N_y$ lattice to the labels of a chain (i.e., a 
one-dimensional array of integer numbers).}
\label{fig:8}
\end{figure}

The mapping just described (indeed it induces an isomorphism of
$*$-algebras)  is the famous Jordan-Wigner transformation
\cite{jordan}. Using this transformation, we can describe any fermionic
unitary evolution in terms of spin operators and therefore simulate
fermionic systems by a quantum computer. Although the mapping as given
is for spinless fermions and for one-dimensional systems, it extends to
higher spatial dimensions and to spin-1/2 fermions by re-mapping each
``mode'' label into a new label corresponding to ``modes'' in a
one-dimensional chain. In other words, if we want to simulate spin-1/2
fermions in a finite $N_x\times N_y$ two-dimensional lattice, we map
the label of the two-dimensional lattice to an integer  number $S$,
running from 1 to $2 (N_x \times N_y)$.   $S$ identifies a mode in the
new chain: 
\begin{eqnarray}
\label{spinfer}
a_{(j,k);\sigma}^{\;} &\rightarrow \tilde{a}_S \rightarrow& 
\left( \prod_{l=1}^{S-1}
-\sigma^l_z \right) \sigma^S_- = (-1)^{S-1} \ \sigma^1_z
\sigma^2_z \cdots \sigma^{S-1}_z \sigma^S_-
\ , \nonumber \\
a^{\dagger}_{(j,k);\sigma} &\rightarrow \tilde{a}_S^{\dagger} \rightarrow& 
\left( \prod_{l=1}^{S-1}
-\sigma^l_z \right) \sigma^S_+ = (-1)^{S-1} \ \sigma^1_z
\sigma^2_z \cdots \sigma^{S-1}_z \sigma^S_+  \ , 
\end{eqnarray}
where the $a_{(j,k);\sigma}$ and $a^{\dagger}_{(j,k);\sigma}$  are the
fermionic spin-1/2 operators in the two-dimensional lattice for the
mode $(j,k)$ and for $z$-component of the spin $\sigma$ ($\sigma=\pm
\frac{1}{2}$), and $\tilde{a}_S$ and  $\tilde{a}_S^{\dagger}$ are the
spinless fermionic operators in the new  chain.  In our case, the modes
are the sites and the label $(j,k)$ identifies  the $X$-$Y$ position of
this site ($j,k \in [1,N_{x,y}]$). The label $(j,k);\sigma$ maps into
the label $S$ (Fig. \ref{fig:8}) via
\begin{equation}
\label{chainmap}
S=j+(k-1)N_x +(\frac{1}{2}-\sigma)N_x N_y \ ,
\end{equation}
This is not the only possible mapping to a two-dimensional lattice
using Pauli matrices \cite{batista01,fradkin,huerta}, but it is very
convenient  for our simulation purposes.

\section{Quantum Network Simulation of a Physical System}
\label{section3}

Like the simulation of a physical system on a classical computer, the
simulation of a physical system on a quantum computer has three basic
steps: the preparation of an initial state, the evolution of the
initial state, and the measurement of the physical properties of the
evolved state. We will consider each process in turn, but first we note
that on a quantum computer there is another important consideration,
namely, the relationship of the operator algebra natural to the
physical system to the algebra of the quantum network. Fortunately, the
mappings (i.e., isomorphisms) between arbitrary representations of 
Lie algebras are now known
\cite{batista01}. Section \ref{section2c} is just one example. To
emphasize this point, the context of our discussion of the three steps
will be the simulation of a system of spinless fermions by the standard
model, which is representable physically as a system of quantum spin
1/2 objects.

\subsection{Preparation of the Initial State} 
\label{section3a}

The preparation of the initial state is important because the
properties we want to measure (correlation functions, energy spectra,
etc.) depend on it. As previously discussed \cite{sfer}, there is a way
to prepare a fermionic initial state of a system with $N_e$ spinless
fermions  and $n$ single particle modes $j$, created by the operators
$a_j^{\dagger}$ ( creation of a fermion in the mode $j$). In the most
general case, the initial state is a linear combination of Slater
determinants 
\begin{equation}
\ket{\Phi_{\alpha}}=\prod\limits_{j=1}^{Ne}b_j^{\dagger} \ \ket{\rm vac}
		   \ ,
\end{equation}
described by the fermionic operators $b_j$ and $b_j^\dagger$, which
are related to the operators $a_j$ and $a_j^{\dagger}$ via a canonical
(unitary) transformation.  Here $\ket{\rm vac}$ is the vacuum state
(zero particle state).  To prepare $\ket{\Phi_{\alpha}}$ one can look
for unitary transformations $U_m$ such that
\begin{equation}
\ket{\Phi_{\alpha}}=e^{i\gamma}\prod\limits_{m=1}^{Ne}U_m \ \ket{\rm vac}
		   \ ,
\end{equation}
where $\gamma$ is a phase factor.  To perform these operations in the standard
model we must express the $U_m$ in terms of Pauli matrices using the
Jordan-Wigner transformation. (We can do the mapping between the Pauli
operators and the $a_{j}$ operators or between the Pauli operators and
the $b_{j}$ operators. In the following we will assume the first
mapping since this will simplify the evolution step.) 
One can choose $U_m = e^{-iH_m t}$ such that
$H_m$ is linear in the $b_m$ and
$b_m^\dagger$ operators~\cite{sfer}. We have to decompose the $U_m$ into
single qubit rotations and two qubit interactions $R_{\mu}
(\vartheta)\mbox{ and }R_{z^{j},z^{k}}(\omega)$. To do this, we first
decompose the $U_{m}$ into a products of operators linear in the $b_m$
or $b_m^{\dagger}$; however, this decomposition does not conserve the
number of particles. The situation appears complex.

Simplification occurs, however, by recalling the Thouless's theorem 
\cite{blaizot} which says that if 
\begin{equation}
\ket{\phi}=\prod\limits_{j=1}^{N_e}a^{\dagger}_j \ \ket{\rm vac}
\end{equation}
and $M$ is a $n\times n$ Hermitian matrix, then
\begin{equation}
e^{i\vec{a}^{\dagger}M\vec{a}} \ \ket{\phi} 
=\prod\limits_{j=1}^{N_e}b^{\dagger}_j \ \ket{\rm vac} \ ,
\end{equation}
where $\vec{a}^\dagger=(a^\dagger_1,\cdots,a^\dagger_n)$ and
\begin{equation}
\label{canonic}
\vec{b}^{\dagger}=e^{iM} \ \vec{a}^{\dagger} \ .
\end{equation}
From Eq. \ref{canonic} the operator $e^{iM}$ (formally acting on the
vector of $a^{\dagger}_j$'s) realizes the canonical transformation
between $a_j$ and $b_j$.

Thouless's theorem generalizes to quantum spin systems via the
Jordan-Wigner transformation. This theorem allows the preparation of an
initial state by simply applying the unitary operator
$e^{i\vec{a}^{\dagger}M\vec{a}}$ to a ``boot up'' state polarized with
each qubit being in the state $\ket{0}$ or $\ket{1}$. Indeed, for an
arbitrary Lie operator algebra the general states prepared in this
fashion are known as Perelomov-Gilmore coherent states
\cite{peregilmo}.

The advantage of this theorem for preparing the initial state instead
of the method previously described \cite{sfer} is that the
decomposition of the unitary operator $e^{i\vec{a}^{\dagger}M\vec{a}}$
can be done in steps, each using combinations of operators $a_{j}
a_{k}^{\dagger}$ and, therefore, conserving the number of particles.
Once the decomposition is done, we then write each operator in terms of
the Pauli operators to build a quantum circuit in the standard model. 
(See Appendix A for a simple example.)

A single Stater determinant is a state of independent particles. That
is, from the particle perspective, it is unentangled.  Generically,
solutions to interacting many-body problems are entangled (correlated)
states, that is, a linear combination of many Slater determinants not
expressible as a single Slater determinant.  In particular, this is
the case if the interactions are strong at short ranges. In quantum
many-body physics, considerable experience and interest exists in
developing simple approaches for generating several specific classes
of correlated wave functions \cite{blaizot}. In Appendix A we
illustrate procedures and recipes to prepare one such class of
correlated (entangled) states, the so-called Jastrow states
\cite{blaizot}.

\subsection{Evolution of Initial State} 
\label{section3b}
The evolution of a quantum state is the second step in the realization 
of a quantum circuit. The goal is to decompose this evolution into the
``elementary gates" $R_{\mu}(\vartheta)\mbox{ and }
R_{z^{j},z^{k}}(\omega)$. To do this for a time-independent
Hamiltonian,  we can write the evolution operator as $\hat{U}(t)=e^{-iHt}$,
where $H=\sum\limits_{l}H_l$ is a sum of individual Hamiltonians $H_l$.
If the commutation relations ${[}H_{l},H_{l'}{]}_{-}=0$ hold for all
$l$ and $l'$, then
\begin{equation}
\hat{U}(t)=\prod\limits_{l}U_{l}(t)=\prod\limits_{l}e^{-iH_{l}t} \ .
\label{Udecomp}
\end{equation}
In this way, we can then decompose each $U_{l}(t)$  in terms of one and two
qubits interactions, using the method described in Section
\ref{section2b}.

In general, the Hamiltonians $H_l$ for different $l$ do
not commute and the relation Eq. \ref{Udecomp} cannot be used. Although
we can in principle exactly decompose the operator $\hat{U}(t)$ into one and two
qubit interactions \cite{barenco:qc1995a,divincenzo:qc1995a},  such a
decomposition is usually very difficult. To avoid this problem, we
decompose the evolution $\hat{U}(t)=\prod\limits_{j}^{\cal M} e^{-iH\Delta
t}$ using the the  first-order Trotter  approximation ($t={\cal M}
\Delta t$):
\begin{equation}
\label{trotter}
\hat{U}(\Delta t)=e^{-iH\Delta t}= e^{-i\sum\limits_l H_l \Delta t } =
\prod_l e^{-i H_l \Delta t} + {\cal O}((\Delta t)^2) \ .
\end{equation}
Then, for $\Delta t \rightarrow 0$, we can approximate the short-time
evolution by: $\hat{U}(\Delta t) \approx \prod\limits_l e^{-i H_l\Delta t }$.
In general, each factor is easily written as one and two qubit
operations (Section \ref{section2b}). 

The disadvantage of this method is that approximating the operator
$\hat{U}(t)$ with high accuracy might require $\Delta t$ to be very small so
the number of steps $e^{-i H_l \Delta t}$ and hence
the number of quantum gates required becomes very large. To mitigate this
problem, we can use a higher-order Trotter decomposition. For example,
if $H=K+V$, we then use the second-order Trotter approximation to
decompose the evolution as $\hat{U}(t) = \prod\limits_{j}e^{-iH\Delta t}$
with (second-order decomposition)
\begin{eqnarray}
e^{-iH\Delta t}&=&e^{-iK \frac{\Delta t}{2}} e^{-iV\Delta t}e^{-iK
\frac{\Delta t}{2}}+{\cal O}((\Delta t)^3) \ , \\
   &=&e^{-iV\frac{\Delta t}{2}}e^{-iK\Delta t}e^{-iV\frac{\Delta t}{2}}+
{\cal O}((\Delta t)^3) \ .
\end{eqnarray}
Other higher-order decompositions are available \cite{suzuki}.

\subsection{Measurement of Physical Quantities}

\subsubsection{One-Ancilla Qubit Measurement Processes}
\label{section3c1}

The last step is the measurement of the physical properties of the
system that we want to study. Often we are interested in measurements
of the form $\langle U^{\dagger}V \rangle$, where $U$ and $V$ are
unitary operators \cite{sfer}. We refer to Ref. \cite{sfer} for a
description of the type of correlation functions that are  related to
these measurements.  See also~\cite{paz} for an application
and variation of these techniques.
Here, we simply give a brief description of how to
perform such measurements.

First, we prepare the system in the initial state $\ket{\Psi_0}$ and
adjoin to it one ancilla (auxiliary) qubit $\sf a$, in the
state $\ket{+}=(\ket{0}+\ket{1})/\sqrt{2}$. This is done by applying
the unitary Hadamard gate to the state $\ket{0}$ (Fig. \ref{fig:4}).
Next, we make two controlled unitary evolutions using the C-$U$ and 
C-$U'$ gates. The first operation  $\tilde{V}$ evolves the system by
$V$ if the ancilla is in the state $\ket{1}$:
$\tilde{V}=\ket{0}\bra{0}\otimes \one+\ket{1}\bra{1}\otimes V$.  The
second one $\tilde{U}$   evolves the system by $U$ if the ancilla state
is $\ket{0}$: $\tilde{U}=\ket{0}\bra{0} \otimes U+\ket{1}\bra{1}\otimes
\one$. ($\tilde{V}$ and $\tilde{U}$ commute.) Once these evolutions
are done, the expectation value of $2\sigma^{\sf a}_{+}= \sigma^{\sf
a}_{x}+i\sigma^{\sf a}_{y}$ gives the desired result. This quantum
circuit is shown in Fig. \ref{fig:9}. Note that the probabilistic
nature of quantum measurements implies that the desired expectation value
is obtained with variance ${\cal O}(1)$ for each instance. 
Repetition can be used to reduce the variance below what is required.

\subsubsection{$L$-Ancilla Qubit Measurement Processes}
\label{section3c2}

Often, we want to compute the expectation value of an operator {\em O} 
of the form
\begin{equation}
O=\sum\limits_{i=1}^Ma_{i} \ U^{\dagger}_{i}V_{i} \ ,
\end{equation}
where $U_i$ and $V_i$ are unitary operators, $a_{i}$ are real positive
numbers ($a_i\geq0$), and $M$ is an integer power of 2. (In the case
that $M$ is less than a power of two, we can complete this definition
by setting the $a_{M+1},\cdots,a_{M'}=0$, where $M'$ is an integer power
of two.) We can compute this expectation value by preparing $M$
different circuits, each one with one ancilla qubit, and for each
circuit measure $\langle U^{\dagger}_iV_i \rangle$ (see
Section~\ref{section3c1}).  Then, we multiply each result by the
constant $a_i$ and sum the results. However, in most cases, the
preparation of the initial state is very difficult. There is another
way to measure this quantity by using only one circuit which reduces
the difficulty.

\begin{figure}
\begin{center}
\includegraphics[angle=0,width=13.0cm,scale=1.0]{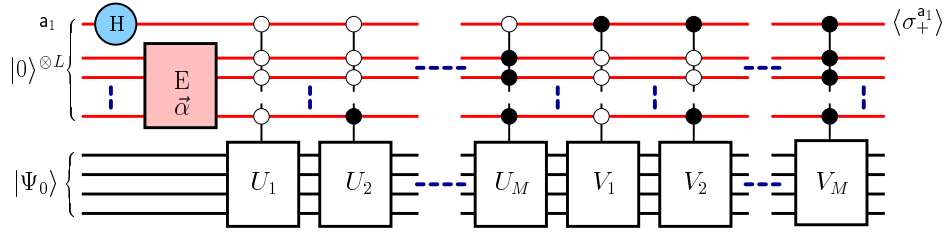}
\end{center}
\bigskip
\caption{Measurement of physical quantities using $L$-ancillas qubits 
$\{{\sf a}_1, \cdots, {\sf a}_L\}$.
In this case $\langle \sigma^{\sf a_1}_+ \rangle= \frac {1} {2 {\cal N}}
\langle \Psi_0|
{[} \sum\limits_{i=1}^M a_i U^{\dagger}_i V_i {]} |\Psi_0 \rangle$ (see
text). }
\label{fig:10}
\end{figure}

We first write the operator $\textit{O}$ as
\begin{equation}
O={\cal N}\sum\limits_{i=1}^M\alpha_{i}^{2} \ U^{\dagger}_{i}V_{i} \ ,
\end{equation}
where ${\cal N}=\sum\limits_{i=1}^Ma_{i}$ and $\alpha_{i}^{2} =
a_{i}/{\cal N}$ ($\sum \limits_{i=1}^M\alpha_{i}^{2}=1$). Then we
construct a quantum circuit with the following steps:
\begin{itemize}
\item[1.]
Prepare the state $\ket{\Psi_0}$ such that $\langle \Psi_0 | O \Psi_0
\rangle$ is the expectation value to be computed.
\item[2.]
Adjoin $L$ ancillas to the initial state, where $L=J+1$ and $2^{J}=M$.
The first of these ancillas, ${\sf a}_{1}$, is prepared in the
state $\ket{+}=(\ket{0}+\ket{1})/ \sqrt{2}$. This is done by applying
the Hadamard gate to the initial state $\ket{0}$ (see Fig.
\ref{fig:4}a).  The other ancillas, $\{{\sf a}_{2},{\sf
a}_{3},\cdots,{\sf a}_{L} \}$ are kept in the state $\ket{0}$.
\item[3.]
Apply a unitary evolution $E(\alpha _1,\alpha _2,\cdots , \alpha _M)$ 
to the ancillas $\{{\sf a}_{2},{\sf a}_{3},\cdots,{\sf a}_{L} \}$ 
to obtain
\[
\ket{\psi}=\alpha_{1}\ket{00\cdots0}+\alpha_{2}
\ket{00\cdots1} +\cdots+\alpha_{M}\ket{11\cdots1}=\sum\limits_{i=1}^M
\alpha_{i} \ \ket{i} \ ,
\]
where $\ket{i}$ is a tensorial product of the states 
($\ket{0}\mbox{ or }\ket{1}$) of each ancilla: 
$\ket{i} = \ket{\eta}_{{\sf a}_2} \otimes \cdots \otimes
\ket{\eta}_{{\sf a}_L}$, where $\eta$ can be 0 or 1. The 
index $i$ orders the orthonormal basis $\ket{i}$.
\item[4.]
Apply the controlled unitary operations $\tilde{U}_{i}$ which
evolve the system by $U_{i}$ if the state of the ancillas is
$\ket{0}_{{\sf a}_1}\ket{i}$. Then apply
the controlled unitary operations $\tilde{V}_{i}$ which evolve the
system by $V_{i}$ if the state of the ancillas is $\ket{1}_{{\sf
a}_1}\ket{i}$. Once these evolution
steps are finished, the state of the whole system is
\[
\ket{\Psi}=\frac{1}{\sqrt{2}} \Biggl[\ket{0}_{{\sf a}_1} \sum
\limits_{i=1}^M \alpha_{i} \ \ket{i} \
U_{i} + \ket{1}_{{\sf a}_1} \sum \limits_{i=1}^M \alpha_{i} \ 
\ket{i} \ V_{i}\Biggr] \otimes \ket{\Psi_0} \ .
\]
\item[5.]
Measure the expectation value of $2 \sigma^{{\sf a}_1}_+=\sigma^{{\sf
a}_1}_x+i\sigma^{{\sf a}_1}_y =2\ket{0}_{{\sf a}_1}\bra{1}$. It is easy
to see that it corresponds to the expectation value of the operator
$\sum\limits_{i=1}^M\alpha_{i}^{2} \ U^{\dagger}_{i}V_{i}$.
\item[6.]
Obtain the expectation value of $\textit{O}$ by multiplying $\langle 2
\sigma^{{\sf a}_1}_+\rangle$ by the constant ${\cal N}$.
\end{itemize}
The quantum circuit for this procedure is given in Fig. \ref{fig:10}.
%\mComment{I found the use of qubit labels inside the ket and state labels
%outside in $\ket{{\sf a}_1}$ confusing and inconsistent with
%the notation $\ket{0}$. Better change that throughout,
%with labels as subscripts and state names inside.}

\subsubsection{Measurement of Correlation Functions}
\label{section3c3}

We now consider measuring correlation functions of the form $C_{AB} =
\langle T^{\dagger} A T B \rangle$, where $T$ is a unitary operator and
$A$ and $B$ are operators that are expressible as a sum of unitary
operators:
\begin{equation}
A=\sum\limits_{i}\alpha _i A_{i} \mbox{ and } B=\sum\limits_{j}\beta _jB_{j}.
\end{equation}
The operator $T$ is fixed by the type of correlation function that we
want to evaluate. In the case of dynamical correlation functions, $T$
is $e^{-iHt}$ where $H$ is the Hamiltonian of the system. For spatial
correlation functions, $T$ is the space translation operator
$e^{-ip\cdot x}$ ($p$ and $x$ are configuration space operators). The
method for measuring these correlation functions is the same method
described in Section ~\ref{section3c1} or Section  ~\ref{section3c2}.
We can use either the one- or the $L$-ancillas measurement process.  

\begin{figure}
\begin{center}
\includegraphics[angle=0,width=13.0cm,scale=1.0]{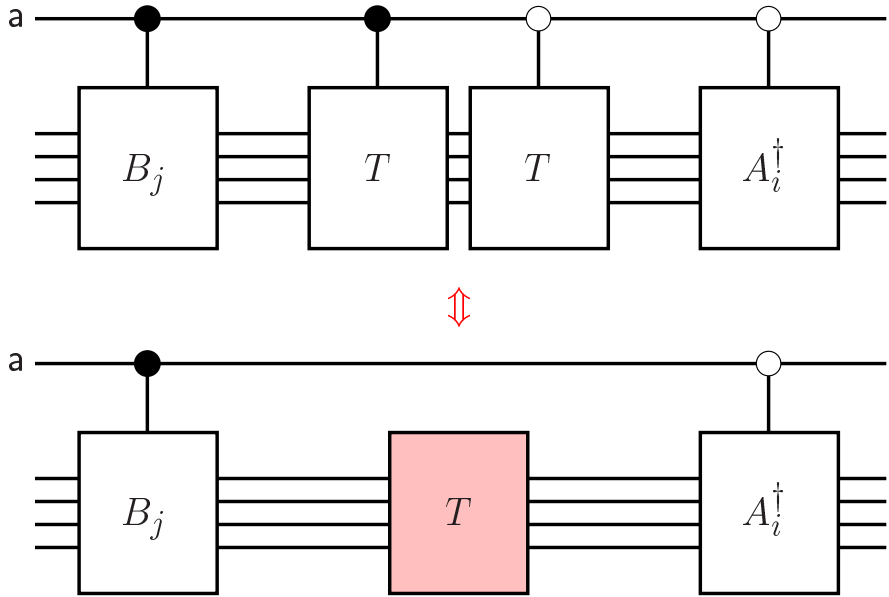}
\end{center}
\bigskip
\caption{Circuit for the measurement of spatial and time correlation functions.
In this case $\langle 2 \sigma_+^{\sf a} \rangle = \langle T^{\dagger} A_i
 T B_j \rangle $. Notice the simplification achieved by reducing two
 C-$T$ operations into only one uncontrolled $T$ operation.}
\label{fig:11}
\end{figure}

To minimize the number of controlled operations and also the quantity
of elementary gates involved, we choose $U^{\dagger}_i=T^{\dagger}A_i$
and $V_j=TB_j$. Now, we have to compute  $\langle U^{\dagger}_iV_j
\rangle$. In Fig. \ref{fig:11} we show the circuit for measuring this
quantity, where the circuit has only one ancilla in the state
$\ket{+}=(\ket{0}+\ket{1})/ \sqrt{2}$. There, the controlled operations
were reduced by noting that the operation of $T$ controlled on the
state $\ket{0}$ of the ancilla followed by the operation of $T$
controlled on the state $\ket{1}$, results in a no-controlled  $T$
operation. This is a very useful algorithmic simplification.

\subsubsection{Measurement of the Spectrum of an Hermitian Operator}
\label{section3c4}

\begin{figure}
\begin{center}
\includegraphics[angle=0,width=13.0cm,scale=1.0]{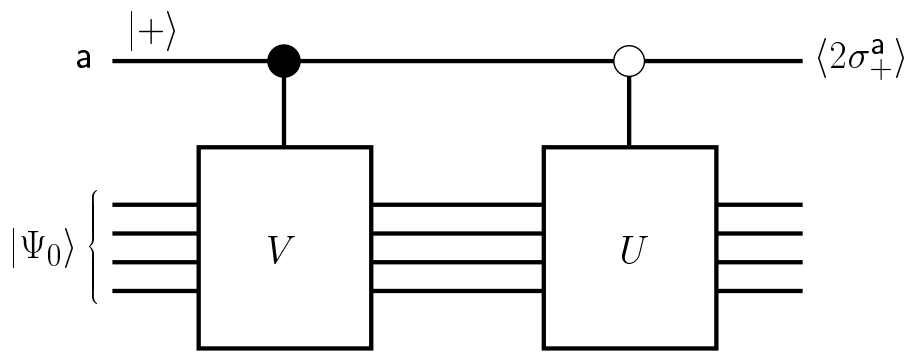}
\end{center}
\bigskip
\caption{Measurement of physical quantities using one extra (ancilla) 
qubit $| {\sf a} \rangle$.
In this case $\langle 2\sigma^{\sf a}_+ \rangle= \langle \Psi_0 | 
U^{\dagger}V \ket{\Psi_0}$.}
\label{fig:9}
\end{figure}

Many times one is interested in determining the spectrum of an
observable (Hermitian operator) $\hat{Q}$, a particular case being the
Hamiltonian $H$. Techniques for getting spectral information can be
based on the quantum Fourier transform~\cite{kitaev,cleve} and can be
applied to physical problems~\cite{abrams}. For our purposes, the
methods of the previous Sections yield much simpler measurements
without loss of spectral information.  For a given $H$, the most
common type of measurement is the computation of its eigenvalues or at
least its lowest eigenvalue (the ground state energy).  To do this we
start from an state $\ket{\phi}$ that has a non-zero overlap with the
eigenstates of $H$.  (For example, if we want to compute the energy of
the ground state, then $\ket{\phi}$ has to have a non-zero overlap
with the ground state.) For finite systems, $\ket{\phi}$ can be the
solution of a mean-field theory (a Slater determinant in the case of
fermions or Perelomov-Gilmore coherent states in the general
case). Once we prepare this state (Section ~\ref{section3a} and
Appendix A), we compute $\langle \hat{U}(t) \rangle=\langle \phi | 
\hat{U}(t) \phi \rangle$, where $\hat{U}$ is the evolution operator 
$\hat{U}(t)=e^{-iHt}$. We then note that
\begin{equation}
\ket{\phi}=\sum\limits_{n=0}^{\cal L} \gamma_{n} \ \ket{\Psi_{n}} \ ,
\end{equation}
with $\ket{\Psi_{n}}$ eigenstates of the Hamiltonian $H$.
Consequently
\begin{equation}
\langle \hat{U}(t) \rangle = \sum\limits_{n=0}^{\cal L} |\gamma_{n}|^{2} \ 
e^{-i{\lambda}_{n}t} \ ,
\label{FFT0}
\end{equation}
where ${\lambda}_{n}$ are the eigenvalues of $H$. The measurement of
$\langle \hat{U}(t) \rangle$ is easily done by the steps described in
Section~\ref{section3c1} (setting $V=\hat{U}(t)$ and $U=\one$ in 
Fig.~\ref{fig:9}). Once we have this
expectation value, we perform a classical  fast Fourier transform 
(i.e., $\int \langle \hat{U}(t) \rangle e^{i\lambda t}dt$) and obtain the
eigenvalues $\lambda_{n}$ (see Appendix~B):
\begin{equation}
\label{FFT}
\mbox{FFT}[\langle \hat{U}(t) \rangle]=\sum\limits_{n=0}^{\cal L}
2 \pi |\gamma_{n}|^{2}{\delta} ({\lambda}-{\lambda}_{n}) \ .
\end{equation}
Although we explained the method for the eigenvalues of $H$, the
extension  to any observable $\hat{Q}$ is straightforward, taking
$\hat{U}(t)=e^{-i\hat{Q}t}$ and proceeding in the same way.

Two comments are in order. The first refers to an algorithmic
optimization and points to decreasing the number of controlled
operations (i.e., the number of elementary gates implemented). If we
set $V=e^{-i\hat{Q}t}$, $U^{\dagger}=\one$ (see Fig.~\ref{fig:9}) 
and perform the type of
measurement described in Section~\ref{section3c1} the network has
total evolution (ancilla plus system) 
$e^{-i\hat{Q}\frac{t}{2}} e^{i\hat{Q}\sigma_z^{\sf a} \frac{t}{2}}$,
while if we set $V=U^{\dagger}=e^{-i\hat{Q} \frac{t}{2}}$
the total evolution is $e^{i\hat{Q}\sigma_z^{\sf a} \frac{t}{2}}$. 
%Then, if we decompose these evolutions into elementary operations
%(assuming that the single qubit rotations $R_{\mu}(\vartheta)$ are
%only about two orthogonal axis $\mu$ and $\mu'$),  the first case
%has $4$ times more operations than the trivial case
%(Section~\ref{section2b}), coming from the difference of complexity
%between the operator $\hat{Q}$ and $\hat{Q}\sigma_z^{\sf a}$. 
Thus, this last algorithm reduces the number of gates by the number of 
gates it takes to represent the operator $e^{-i\hat{Q}\frac{t}{2}}$. 
The circuit is shown in Fig.~\ref{fig:12}.

The second comment refers to the complexity of the quantum algorithm as
measured by system size. In general it is difficult to find a state
whose overlap scales polynomially with system size. If one chooses a
mean-field solution as the initial state, then the overlap decreases
exponentially with the system size; this is a ``signal problem'' which
also arises in probabilistic classical simulations of quantum systems.
The argument goes as follows: If $|\phi\rangle$ is a mean-field
state for an $N^d$(=volume) system size whose (modulus of the) 
overlap with the true eigenstate is $|\gamma|<1$, and assuming 
that the typical correlation length of the problem $\xi$ is smaller 
than the linear dimension $N$, if we double $N$ the new overlap is 
$\sim e^{2^d \ln |\gamma|}$.

We would like to mention that an alternative way
of computing part of the spectrum of an Hermitian operator is using the
adiabatic connection or Gell-Mann-Low theorem, an approach that has 
been described in \cite{sfer}. 

\subsubsection{Mixed and Exact Estimators}

We already explained how to compute different types of correlation
functions.  But in most  cases, we do not know the state whose
correlations we want to obtain. The most common case is wanting the
correlations in the ground state $\ket{\Psi_0}$ of some Hamiltonian
$H$.  Obtaining the ground state is a very difficult task; however,
there are some useful methods to approximate these correlation
functions.

Suppose we are interested in the mean value of a unitary operator
$O(t)$.  If we can prepare the initial state $\ket{\Psi_T}$ in such a
way that $\ket{\Psi_0}=\ket{\Psi_T}+\epsilon \ket{\Phi}$ ($\epsilon$
is  intended to be small), then after some algebraic
manipulations \cite{negele:qc1988a}, we have
\begin{equation}
\label{mixest}
\frac{\bra{\Psi_0}O(t)\ket{\Psi_T}} {\langle \Psi_0 | \Psi_T\rangle}
=\frac{1} {2} \left [
\frac{\bra{\Psi_0}O(t)\ket{\Psi_0}} {\langle \Psi_0 | \Psi_0 \rangle} +
\frac{\bra{\Psi_T}O(t)\ket{\Psi_T}} {\langle \Psi_T | \Psi_T \rangle}  
\right ] + 
{\cal O} (\epsilon^2)
\end{equation} 
where the term on the left-hand side of Eq.~\ref{mixest} is known as
the ``mixed estimator.''  Also, we can calculate the second term on the
right-hand side of Eq.~\ref {mixest} with an efficient quantum 
algorithm, since we are able to prepare easily $\ket{\Psi_T}$. Next, we
show how to determine the mixed estimator using a quantum algorithm.

If $\ket{\Psi_0}$ is the ground state, then it is an eigenstate of the
evolution  operator  $\hat{U}(t')=e^{-iHt'}$, and we can obtain the mixed
estimator by measuring the mean  value of $\hat{U}(t')O(t)$: Because
$\ket{\Psi_T}= \sum\limits_n a_n\ket{\Psi_n}$ where $a_n=\langle\Psi_n|
\Psi_T\rangle$ and $\ket{\Psi_n}$ are the eigenstates of $H$
($\hat{U}(t')\ket{\Psi_n}=e^{-i\lambda_nt'} \ket{\Psi_n}$) we can measure
(Section ~\ref{section3c3})
\begin{equation}
\label{mixest2}
\bra{\Psi_T}\hat{U}(t')O(t)\ket{\Psi_T}=\sum\limits_n e^{i\lambda_nt'}
\langle\Psi_T |\Psi_n\rangle \bra{\Psi_n}O(t)\ket{\Psi_T}
\end{equation}
By performing a Fourier transform in the variable $t'$
($\tilde{F}(\omega)= \int e^{i\omega t'}F(t')dt'$) in Eq.~\ref
{mixest2} and making the relation  between the expectation value for
time $t$ and the expectation value for $O(t)=\one$, we obtain the value
of the mixed estimator.  Then, by using Eq.~\ref {mixest}, we obtain
$\frac{\bra{\Psi_0}O(t)\ket{\Psi_0}} {\langle \Psi_0 | \Psi_0 \rangle}$
up to order $\epsilon^2$.

By similar steps, we can obtain expectation values of the form $\frac
{\bra{\Psi_n}O(t)\ket{\Psi_{n'}}} {\langle \Psi_n | \Psi_{n'} \rangle}$
for all $n$ and $n'$. The trick consists of measuring (Section
~\ref{section3c3}) the mean value of the  operator
$\hat{U}(t')O(t)\hat{U}^{\dagger}(t'')$ in the state $\ket{\Psi_T}$
\begin{equation}
\bra{\Psi_T}\hat{U}(t')O(t)\hat{U}^{\dagger}(t'')\ket{\Psi_T}=
\sum\limits_{n,n'} e^{i\lambda_n t'}e^{i\lambda_{n'} t''}
\langle \Psi_T | \Psi_n \rangle \langle \Psi_{n'} | \Psi_T \rangle
\bra{\Psi_n} O(t) \ket{\Psi_{n'}}
\end{equation}
and then by performing a double Fourier transform in the variables 
$t'$ and $t''$ ($\tilde{F}(\lambda,\lambda ')=\int e^{i\lambda
t'}e^{i\lambda' t''}F(t',t'') dt'dt''$) we obtain the desired results.
A particular case of this  procedure is the direct computation of the
exact estimator  $\frac{\bra{\Psi_n}O(t)\ket{\Psi_n}} {\langle \Psi_n |
\Psi_n \rangle}$.

\section{Application to Fermionic Lattice Systems}
\label{section4}

In this Section, we illustrate a procedure for simulating fermionic
systems on a quantum computer, showing as a particular example how to
obtain the energy spectrum of the Hubbard Hamiltonian for a
finite-sized system. We will obtain this spectrum through a simulation
of a quantum computer on a classical computer, that is, by a quantum
simulator.

We start by noting that the spin-fermion connection described in
Eq.~\ref{spinfer} and Eq.~\ref{chainmap} implies that the number of
qubits involved in a two-dimensional lattice is $L=2(N_x\times N_y)$ if
one uses the standard model to simulate spin-1/2 fermions.  Also, the
number of states for an $L$-qubits system is $2^L$. From this mapping,
the first $N_x\times N_y$ qubits represent the states which have
spin-up fermions, and the other qubits ($(N_x\times N_y+1)$ to
$2(N_x\times N_y)$) spin-down fermions.  In other words, if we have a
system of 4 sites and have a state $\ket{\Psi}$ with one electron with
spin up at the first site and one electron with spin down at the third
site, then this state in second quantization is $\ket{\Psi} =
a_{1;\uparrow}^{\dagger}a_{3;\downarrow}^{\dagger} \ket{\mbox{vac}}$,
where the fermionic operator $a_{j;\sigma}^{\dagger}$ creates a fermion
in the site $j$ with spin $\sigma$, and $\ket{\mbox{vac}}$ is the state
with no particles (vacuum state). In the standard model, this state
corresponds to
\begin{equation}
\ket{\Psi}=\sigma_+^1 \prod_{l=1}^6\sigma_z^l\sigma_+^7
\ket{\widetilde{\mbox{vac}}}
=\ket{0}\otimes\ket{1}\otimes\ket{1}\otimes\ket{1}\otimes\ket{1}\otimes
\ket{1}\otimes\ket{0}\otimes\ket{1}\rightarrow
\ket{\uparrow\downarrow\downarrow\downarrow
\downarrow\downarrow\uparrow\downarrow} ,
\label{stateexample}\end{equation}
where $\ket{\widetilde{\mbox{vac}}}$ is the vacuum of the quantum spin
1/2, which we have chosen to be $\ket{\downarrow\downarrow\downarrow
\downarrow\downarrow\downarrow\downarrow\downarrow}$.
 
\begin{figure}
\begin{center}
\includegraphics[angle=0,width=13.0cm,scale=1.0]{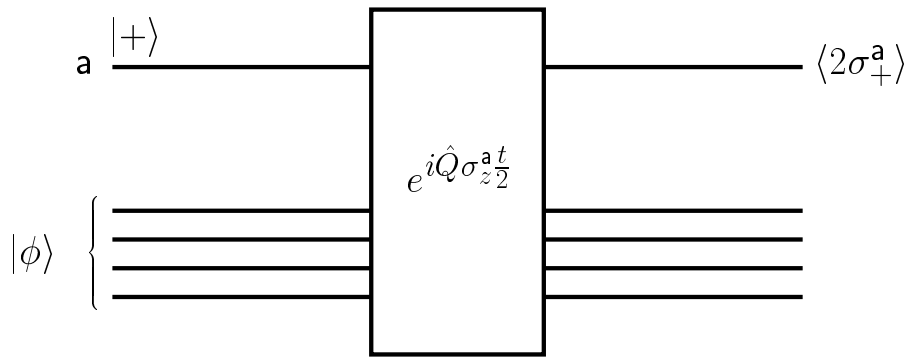}
\end{center}
\bigskip
\caption{Circuit for the measurement of the spectrum of an Hermitian operator
$\hat{Q}$. In this case $\langle 2\sigma^{\sf a}_+ \rangle = \langle \phi
| e^{-i \hat{Q} t} \phi \rangle $ (see text).}
\label{fig:12}
\end{figure}

To represent the $L$-qubit system on a classical computer, we can build
a one-to-one mapping between the $2^L$ possible states and the bit
representation of an integer $I$ defined by
\begin{equation}
\label {binamap}
I=\sum\limits_{i=1}^L [n(i)\times 2^{i-1}]
\end{equation}
where $n(i)$ (occupancy) is 0 if the spin of the $i$-qubit is 
$\ket{1}$ ($\downarrow$), or 1 if the state is $\ket{0}$ ($\uparrow$).
In this way, the state described in Eq.~\ref{stateexample} maps to
$I=65$. Because we are interested in obtaining some of the eigenvalues
of the Hubbard model, we added an ancilla qubit (Fig.~\ref{fig:12}).
The ``new'' system has $L=2(N_x \times N_y)+1$ qubits, and we can
perform the mapping in the same  way described above.

To simulate the evolution operator $\hat{U}(t)=e^{-iHt}$ on a classical
computer using the above representation of quantum states, we
programmed the ``elementary" quantum gates of one and two qubits
interactions.  Each $L$-qubit state was represented by a linear
combination of the integers $I$ (Eq.~\ref{binamap}).  In this way,
each unitary operation applied to one or two qubits modifies $I$ by
changing a bit. For example, if we flip the spin of the first qubit,
the number $I$ changes by 1.

\begin{figure}
\begin{center}
\includegraphics[angle=0,width=13.0cm,scale=1.0]{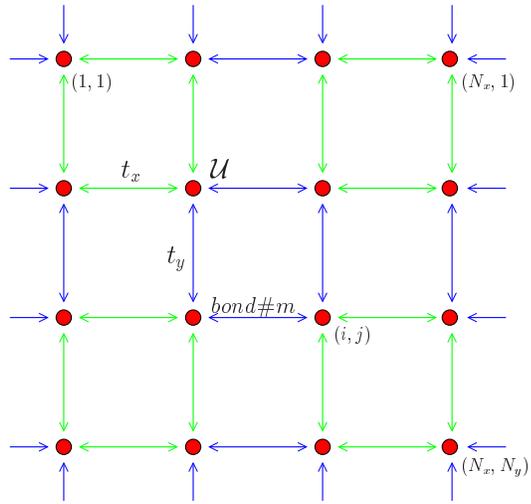}
\end{center}
\bigskip
\caption{Two-dimensional lattice in the Hubbard model. Here, the green and blue
arrows identify the even and odd bonds.}
\label{fig:13}
\end{figure}

We want to evaluate some eigenvalues of the spin-1/2 Hubbard model in
two spatial dimensions. The model is defined on a rectangle of
$N_x\times N_y$ sites and is parametrized by spin preserving hoppings
$t_x$ and $t_y$ between nearest neighbor sites, and an interaction
$\cal U$ on site between fermions of different $z$-component of spin
(Fig.~\ref{fig:13}). The Hamiltonian is
\begin{eqnarray}
H=-\sum\limits_{(i,j);\sigma}
[&&\!\!\!\!\!\! t_x(a^{\dagger}_{(i,j);\sigma}a_{(i+1,j);\sigma}^{\;}
     +a^{\dagger}_{(i+1,j);\sigma}a_{(i,j);\sigma})^{\;}
  + t_y(a^{\dagger}_{(i,j);\sigma}a_{(i,j+1);\sigma}^{\;}
    +a^{\dagger}_{(i,j+1);\sigma}a_{(i,j);\sigma}^{\;})]\nonumber\\
&+&{\cal U}\sum\limits_{(i,j)}n_{(i,j);\uparrow}n_{(i,j);\downarrow} \ ,
\end{eqnarray}
where $n_{(i,j);\sigma} = a^{\dagger}_{(i,j);\sigma}
a_{(i,j);\sigma}^{\;}$ is the number operator, and the label
$(i,j);\sigma$ identifies  the site ($X$-$Y$ position) and the
$z$-component of spin ($\sigma=\pm 1/2$). We assume the fermionic
operators satisfy strict periodic boundary conditions in both
directions: $a_{(i,j);\sigma}=a_{(i+N_x,j);\sigma}=
a_{(i,j+N_y);\sigma}$.

To obtain the energy spectrum for this model, we use the method
described in Section~\ref{section3c3}. (See Fig.~\ref{fig:12}.)  
For this, we represent the system in the standard model, using the
Jordan-Wigner transformation, mapping a two-dimensional
spin-1/2 system into a one-dimensional chain, with the use
of Eq.~\ref{chainmap} and Eq.~\ref{spinfer} (Fig.~\ref{fig:8}).

As explained in Section~\ref{section3c4}, we find it convenient to
start from the mean-field ground state solution of the model,
represented by $H_{MF}$
\begin{eqnarray*}
H_{MF}=-\sum\limits_{(i,j);\sigma}
[&&\!\!\!\!\!\! t_x(a^{\dagger}_{(i,j);\sigma}a_{(i+1,j);\sigma}^{\;}
        +a^{\dagger}_{(i+1,j);\sigma}a_{(i,j);\sigma}^{\;})
  + t_y(a^{\dagger}_{(i,j);\sigma}a_{(i,j+1);\sigma}^{\;}
        +a^{\dagger}_{(i,j+1);\sigma}a_{(i,j);\sigma}^{\;})]\\
&+&{\cal U}\sum\limits_{(i,j)}
[\langle n_{(i,j);\uparrow}\rangle n_{(i,j);\downarrow}
 +n_{(i,j);\uparrow}\langle n_{(i,j);\downarrow}\rangle 
 -\langle n_{(i,j);\uparrow} \rangle \langle n_{(i,j);\downarrow}
 \rangle] \ ,
\end{eqnarray*}
where the expressions in angular brackets are expectation values in the
mean-field representation. Without loss of generality, we take ${\cal
U}>0$ and select the anti-ferromagnetic ground state mean-field
solution.  For this solution, we require $N_x$ and $N_y$ to be even
numbers. If we were to simulate a one-dimensional lattice, we would
however chose one of these numbers to be even and the other equal to 1.
In the following we will only consider the half-filled case which 
corresponds to having  one fermion per site; i.e., $N_e=N_x \times N_y$).

First, we prepare the initial state.  As discussed in Section
~\ref{section3a}, we do this by exploiting Thouless's theorem. We also
use the first-order Trotter approximation (Section ~\ref{section3b}),
and then decompose each term of the evolution into one and two qubit
interactions.  Here, the matrix $M$ now depends on the parameters of
the Hamiltonian, as it does the ground state mean-field solution. 
After the decomposition, we then prepare the desired initial state by
applying the unitary evolutions to a polarized state. (See Appendix A).

Next, we execute the evolution $\hat{U}(t)=e^{-iHt}$. For the sake of
clarity we only present the first-order Trotter decomposition.  To
this end, we rewrote the Hubbard Hamiltonian as
\begin{equation}
H=K+V=K_{\uparrow}+K_{\downarrow}+V \ ,
\end{equation}
where $K_{\sigma}$ is the kinetic term (hopping elements with spin
${\sigma}$) and $V$ is the potential energy term.  Because
${[}K_{\sigma},V{]}_-\neq 0$ and ${[}K_{\uparrow},
K_{\downarrow}{]}_-=0$ we approximated the short-time evolution
operator $\hat{U}(\Delta t)$ by
\begin{equation}
\hat{U}(\Delta t)=e^{-iH\Delta t}\approx e^{-iK\Delta t}e^{-iV\Delta t}\mbox{;  }
(\Delta t\rightarrow 0) \ .
\end{equation}
Because the term $V={\cal U}\sum\limits_{(i,j)}n_{(i,j);\uparrow}
n_{(i,j);\downarrow}  =\sum\limits_{l=1}^{N_x\times N_y}V_l$ is a sum
of operators local to each lattice site, each of these terms commute so
\begin{equation}
e^{-iV\Delta t}=\prod\limits_{l}e^{-iV_{l}\Delta t} \ .
\end{equation}

The kinetic term is a sum over the bonds in the lattice
(Fig.~\ref{fig:13}): $K_{\sigma}=\sum\limits_{bonds}K_{bond;\sigma}$.
Each bond joins two nearest neighbor sites, either in the vertical or
horizontal direction (Fig. \ref{fig:13}).  Because of the periodic
boundary conditions, the sites at the boundary of the lattice are also
connected by bonds.  We note that the terms in $K$ that share a lattice
site do not commute. For these terms we rewrite $K_{\sigma}$ as
\begin{equation}
\label{bondes}
K_{\sigma}=K_{x;\sigma}^o+K_{x;\sigma}^e+K_{y;\sigma}^o+K_{y;\sigma}^e \
,
\end{equation}
where $K_{\mu ;\sigma}^{e(o)}$ are the kinetic terms (for spin
$\sigma$) in the $\mu$-direction that involve the even ($e$)  (and odd
($o$)) bonds in this direction (green and blue lines in
Fig.~\ref{fig:13}). Then we perform the first-order Trotter
approximation
\begin{equation}
e^{-iK_{\sigma}\Delta t}\approx e^{-iK_{x;\sigma}^o\Delta t}
e^{-iK_{x;\sigma}^e \Delta t}e^{-iK_{y;\sigma}^o\Delta
t}e^{-iK_{y;\sigma}^e\Delta t} \ .
\end{equation}

Because the odd and even bonds are not connected, each term in
(\ref{bondes}) is a sum of terms that commute with each other, that is:
$K_{\mu ;\sigma}^{e(o)}= \sum\limits_m K_{\mu ;\sigma}^{e(o);m}$,
where  ${[}K_{\mu ;\sigma}^{e(o);m},K_{\mu ;\sigma}^{e(o);m'}{]}_- =0$,
then:
\begin{equation}
e^{-iK_{\mu ;\sigma}^{e(o)}\Delta t}=
\prod\limits_m e^{-iK_{\mu ;\sigma}^{e(o);m}\Delta t}  \ .
\end{equation}

In summary we approximated the short-time evolution $\hat{U}(\Delta t)$ by
\begin{equation}
\hat{U}(\Delta t)\approx {\Bigl[}\prod\limits_{m_1,m_2,m_3,m_4;\sigma}
e^{-iK_{x;\sigma}^{o;m_1}\Delta t}e^{-iK_{x;\sigma}^{e;m_2}\Delta t}
e^{-iK_{y;\sigma}^{o;m_3}\Delta t}e^{-iK_{y;\sigma}^{e;m_4}\Delta t}{\Bigr]}
{\Bigl[}\prod\limits_{l}e^{-iV_{l}\Delta t}{\Bigr]} \ .
\end{equation}
The total evolution operator is
\begin{equation}
\hat{U}(t)=\prod_j \hat{U}(\Delta t) \ .
\end{equation}
Each unitary factor in the evolution is easily decomposed into
one and two qubit interactions (Section~\ref{section2b}). 

The final step is the measurement process. To obtain some of the
eigenvalues, we use the circuit described in Fig.~\ref{fig:12}. Thus we
are interested in the operator $\hat{U}(t/2)^{-\sigma_z^{\sf a}}$ instead of
$\hat{U}(t/2)$ so we actually performed the first two steps after adding an
ancilla qubit ${\sf a}$ (Fig.~\ref{fig:12}), and then started with a
``new'' Hamiltonian $\tilde{H}=-H\otimes \frac{\sigma_z^{\sf a}}{2}$,
(and also a ``new'' evolution $\tilde{U}(t)=e^{-i\tilde{H}t}$) and
performed the same steps described above.

The results for the simulation of the Hubbard model are shown in
Fig.~\ref{fig:14}. There, we also show the parameters $\Delta t_1$ and
$\Delta t_2$ corresponding to the time-steps we used in the initial
state preparation, where we used a first-order Trotter approximation,
and in the time evolution, where we used a second-order Trotter
approximation.

In closing this Section we would like to emphasize that the simulation
of the Hubbard model by a quantum computer which uses the standard
model is just an example. Suppose one wants to simulate the Anderson
model \cite{bonca} instead using the same quantum computer, then
similar steps  to the  ones described above should be followed. (There
are two types of fermions but the isomorphism still applies.)
Similarly, if one wants to use a different quantum computer which has
another natural ``language'' (i.e., a different operator algebra which
therefore represents a different model of computation) one can still
apply the ideas developed above simply by choosing the right
isomorphism or ``dictionary'' \cite{batista01}.

\section{Concluding Remarks}

We addressed several broad issues associated with the simulation of
physical phenomena by quantum networks. We first noted that in quantum
mechanics the physical systems we want to simulate are expressed by
operators satisfying certain algebras that may differ from the
operators and the algebras associated with the physical system
representing the quantum network used to do the simulation. We pointed
out that rigorous mappings \cite{batista01} between these two sets of 
operators exist and are sufficient to establish the equivalence of the
different physical models to a universal model of quantum computation
and the equivalence of different physical systems to that model.

We also remarked that these mappings are insufficient for establishing
that the quantum network can simulate any physical system efficiently
even if the mappings between the systems only involves a polynomial
number of steps. We argued that one must also demonstrate the main
steps of initialization, evolution, and measurement all scale
polynomially with complexity. More is needed than just having a large
Hilbert space and inherent parallelism. Further, we noted that some
types of measurements important to understanding physical phenomena
lack effective quantum algorithms.

In this paper we mainly explored various issues associated with
efficient physical simulations by a quantum network, focusing on the
construction of quantum network models for such computations. The main
questions we addressed were how do we reduce the number of qubits and 
quantum gates needed for the simulation and how do we increase the
amount of physical information measurable. We first summarized the
quantum network representation of the standard model of quantum
computation, discussing both one and multi-qubit circuits, and then
recalled the connection between the spin and fermion representations. We
next discussed the initialization, evolution, and measurement
processes. In each case we defined procedures simpler than the ones
presented in our previous paper \cite{sfer}, greatly improving the 
efficiency with
which they can be done. We also gave algorithms that greatly expanded
the types of measurements now possible. For example, besides certain
correlation functions, the spectrum of operators, including the energy
operator, is now possible. Our application of this technology to a
system of lattice fermions and the construction of a simulator was also
discussed and used the Hubbard model as an example. This application
gave an explicit example of how the mapping between the operator of the
physical system of interest and those of the standard model of quantum
computation work. We also gave details of how we implemented the
initialization, evolution, and measurement steps of the quantum network
on a classical computer, thereby creating an quantum network simulator.

Clearly, a number of challenges for the efficient simulation of
physical systems on a quantum network remain. We are prioritizing our
research on those issues associated with problems that are 
extremely difficult for 
quantum many-body scientists to solve on classical computers.  There
are no known efficient quantum algorithms for broad spectrum
ground-state (zero temperature) and thermodynamics (finite
temperature) measurements of correlations in quantum states. These
measurements would help establish the phases of those
states. Generating those states is itself a difficult task. 

Many problems in physics simulation, such as the challenging
protein folding problem, are considered to be well modeled
by classical physics. Can quantum networks be used to obtain
significantly better (more efficient) algorithms for such 
essentially classical physics problems?

\begin{acknowledgments}
We thank Ivar Martin for useful discussions on the classical Fourier
transform. 
\end{acknowledgments}

% APPENDICES
\newpage
\appendix
\section{Different state preparation}
\subsection{Coherent State Preparation: An example}

Here we illustrate by example the decomposition of an operator of
the form $e^{i \vec a^\dagger M\vec a}$ to generate an initial
state. Typically $M$ is generated by some mean-field solution to the 
physical problem of interest.  Considerable detail is given.

We consider 2 spinless fermions in a one-dimensional lattice of 4 sites
($N_{e}=2$ , $n=4$). The operators $a_j$ and $a^{\dagger}_j$ annihilate
and create a fermion in the site $j$ of the lattice. We want to prepare
an initial state $\ket{\phi '}=c^{\dagger}_{0} c^{\dagger}_{\pi
/2}\ket{\mbox{vac}}$ from the state $\ket{\phi }=a^{\dagger}_{1}
a^{\dagger}_{2}\ket{\mbox{vac}}$, where the operators $c_k$ and
$c^{\dagger}_{k}$ annihilate and create a fermion in the state of wave
vector $k$, that is:
\begin{equation}
\label{wavemap}
c^{\dagger}_{k}=\frac{1}{2}\sum\limits_{j=1}^{4}e^{ikx_j}a^{\dagger}_{j}
\ ,
\end{equation}
where $k=0,{\pi /2},{\pi},{3\pi /2}$ are all possible wave vectors of
the system, and $x_j$ is the position in the lattice of the site
(i.e., $x_j=j-1$).

From Eq. \ref{wavemap}, we see that the state $\ket{\phi '}$ is a
linear combination of states of the form
$a^{\dagger}_{i}a^{\dagger}_{j}\ket{\mbox{vac}}$. The change of basis
$e^{iM}$ (Eq.~\ref{canonic}) between the two sets of fermionic
operators is:
\begin{equation}
\label{chbasis}
\pmatrix{c_0^{\dagger} \cr c_{\pi /2}^{\dagger}\cr
c_{\pi}^{\dagger} \cr c_{3\pi /2}^{\dagger} \cr} = \frac{1}{2} \
\pmatrix{ 1 & 1 & 1 & 1\cr 1 & i & -1 & -i \cr 1 & -1 & 1 & -1 \cr
1 & -i & -1 & i \cr} \ \pmatrix {a_1^{\dagger} \cr
a_{2}^{\dagger}\cr a_{3}^{\dagger} \cr a_{4}^{\dagger} \cr} \ .
\end{equation}
If we calculate the eigenvalues and the eigenvectors of the matrix
$e^{iM}$, from Eq.~\ref{chbasis} we obtain:
\begin{equation}
e^{iM_D}=\ \pmatrix{ -1 & 0 & 0 & 0\cr 0 & i & 0 & 0 \cr
0 & 0 & 1 & 0 \cr 0 & 0 & 0 & 1 \cr} \ ,
\end{equation}
where $M_D$ is $M$ in its diagonal form. Then, we have:
\begin{equation}
M_D=-i \log (e^{iM_D})=\ \pmatrix{ \pi & 0 & 0 & 0\cr 0 & \pi /2 &
0 & 0 \cr 0 & 0 & 0 & 0 \cr 0 & 0 & 0 & 0 \cr} \ .
\end{equation}
To obtain the matrix $M=A^{\dagger}M_{D}A$, we need to know the
unitary matrix $A$, which is constructed with the eigenvectors of
the matrix $e^{iM}$. In this case we have:
\begin{equation}
A^{\dagger}=\ \pmatrix{ -1/2 & 0 & 1/\sqrt{2} & 1/\sqrt{2}\cr 1/2
& -1/\sqrt{2} & 1/\sqrt{2} & 0 \cr 1/2 & 0 & -1/\sqrt{2} &
1/\sqrt{2} \cr 1/2 & 1/\sqrt{2} & 1/\sqrt{2} & 0 \cr} \ ,
\end{equation}
hence, the Hermitian matrix $M$ is:
\begin{equation}
M=\frac{\pi}{4} \ \pmatrix{ 1 & -1 & -1 & -1\cr -1 & 2 & 1 & 0 \cr
-1 & 1 & 1 & 1 \cr -1 & 0 & 1 & 2 \cr} \ .
\end{equation}

In order to obtain $\ket{\phi '}$ we prepare the state $\ket{\phi}$ and
then apply the evolution  $U=e^{i\vec{a}^{\dagger} M \vec{a}}$. If we
want to simulate this fermionic system in a quantum computer (standard
model),  we have to use the spin-fermion connection
(Section~\ref{section2c}), and write the operator $U$ as a combination
of single qubit rotations and two qubit interactions. Also, the initial
state $\ket{\phi}$ must be written in the standard model:
\begin{eqnarray}
\ket{\phi}&=&a^{\dagger}_{1}a^{\dagger}_{2}\ket{\mbox{vac}}=
\sigma^1_+(-\sigma^1_z\sigma^2_+) \ket{\widetilde{\mbox{vac}}}
           =\sigma^1_+ \sigma^2_+\ket{\widetilde{\mbox{vac}}} \\
          &=&\ket{0}_1\otimes\ket{0}_2\otimes\ket{1}_3\otimes\ket{1}_4
	  = \ket{\uparrow\uparrow\downarrow\downarrow} \ ,
\label{inistate}\end{eqnarray}
where the vacuum state in the standard model is
$\ket{\widetilde{\mbox{vac}}}=\ket{1}_1\otimes
\ket{1}_2\otimes\cdots\otimes\ket{1}_n=\ket{\downarrow\downarrow
\cdots  \downarrow}$ ($(\prod\limits_{l=1}^
{j-1}-{\sigma}_z^l)\sigma_-^j\ket{\widetilde{\mbox{vac}}} =
a_j\ket{\mbox{vac}}=0$). With this mapping, the state $\ket{\phi '}$
is a linear combination of states of $z$-component of spin 0.

\begin{figure}
\begin{center}
\includegraphics[angle=0,width=13.0cm,scale=1.0]{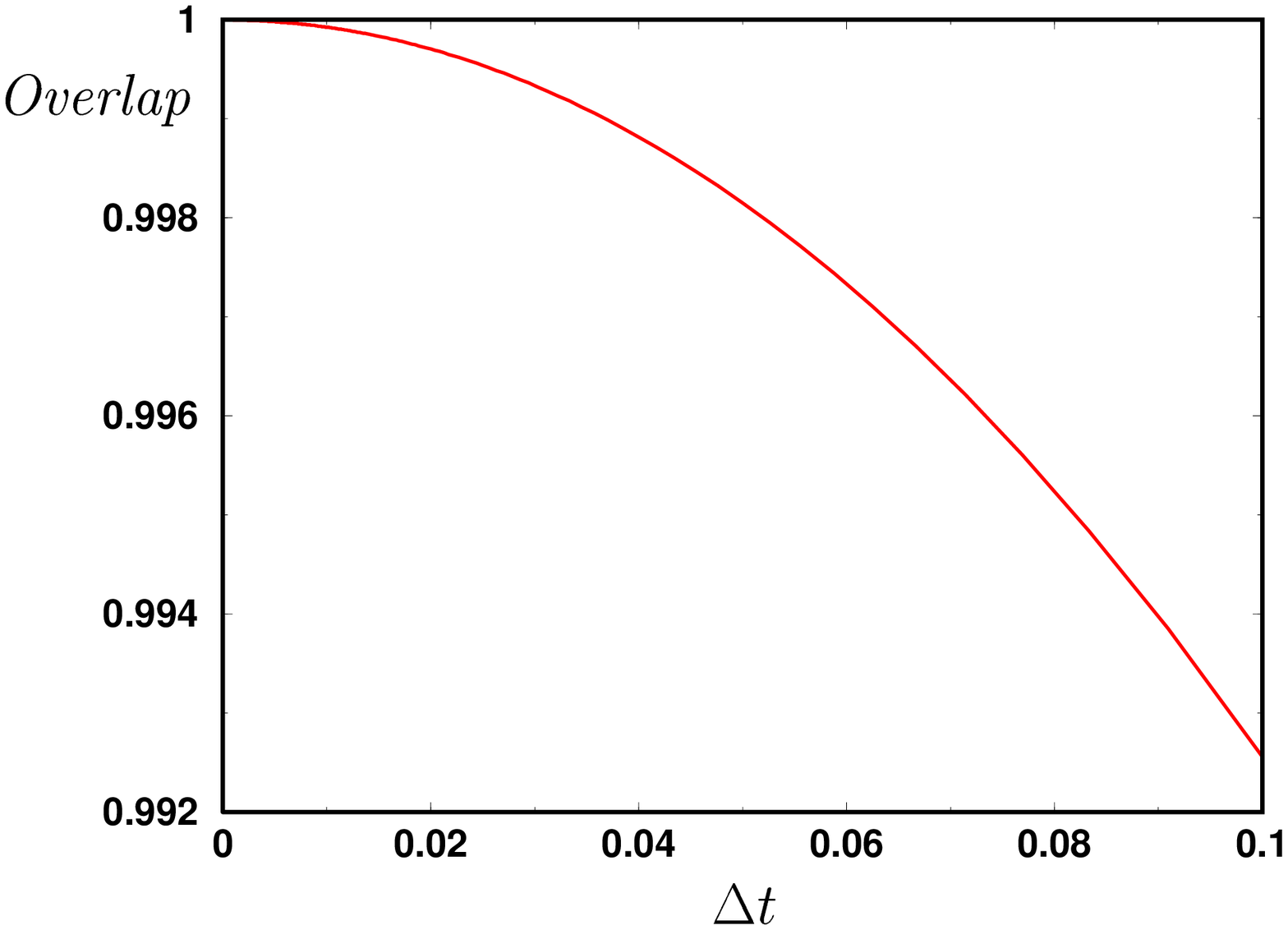}
\end{center}
\bigskip
\caption{Overlap between the exact initial state and the state prepared
with the Trotter decomposition, for a system with two fermions in a 4-sites
lattice.}
\label{fig:15}
\end{figure}

As noted in Section~\ref{section3b},  sometimes the decomposition
of the operator $U$  in terms of one and two qubit operations is very
difficult. To avoid this problem, we can use the Trotter decomposition
(Eq. \ref{trotter}). In Fig.~\ref{fig:15} we show the overlap
(projection) between the state  $\ket{\phi '}$ and the state prepared
using the first-order Trotter decomposition of $U$ applied to the state
$\ket{\phi}$.

\subsection{Jastrow-type Wave Functions}

A Jastrow-type wave function is often a better approximation to the
actual state of an interacting system, particularly when interactions
are strong and short-ranged. Often one varies the parameters in these 
functions to produce a state that satisfies a variational principle
for  some physical quantity like the energy. Such states build in
correlated  many-body effects and are, in general, entangled states.
The states  described in the previous subsection (Appendix A 1) are
unentangled.

The classic form of a Jastrow-type wave function for fermions
is\cite{blaizot}
\begin{equation}
\ket{\Psi_0}=e^S \ket{\phi'} \ ,
\end{equation}
where $S=\sum\limits_{ij} \alpha_{ij} c^{\dagger}_i
c_j^{\;}+\sum\limits_{ijkl} \beta_{ijkl} c^{\dagger}_i c^{\dagger}_j
c_k^{\;} c_l^{\;} + \cdots$  is an operator which creates particle and
hole excitations,  and $\ket{\phi'}$ is typically a Slater determinant.
The $N$-body  correlations embodied in $S$ take into account the
short-range forces  not included in $\ket{\phi'}$. We will assume the
$\alpha_{ij}$ and $\beta_{ijkl}$ have been determined by some suitable
means (for example, by a coupled-cluster calculation).  If we decompose
$e^S$ into a linear combination of unitary operators, we can then
decompose $\ket{\Psi_0}$ into a linear combination of Slater
determinants and thus prepare $\ket{\Psi_0}$ as explained in
\cite{sfer}.  Also, if the coefficients $\alpha_{ij}$ and
$\beta_{ijkl}$ are small, we can approximate $e^S$ by the first few
terms in its Taylor expansion.  Again, the state $\ket{\Psi_0}$ will be
a linear combination of Slater determinants. 

Obviously, it is more natural for a quantum computer to generate a 
correlated state of the form 
\begin{equation}
\ket{\Psi_0}=e^{i S} \ket{\phi'} \ ,
\end{equation}
where $e^{i S}$ is a unitary operator. In order to determine the
$N$-body correlation coefficients $\alpha_{ij}$ and $\beta_{ijkl}$,  
one could, in principle, use the technique of unitary transformations 
introduced by Villars \cite{villars}. 

\newpage

\section {Discrete Fourier Transforms}

\begin{figure}
\begin{center}
\includegraphics[angle=270,width=13.0cm,scale=1.0]{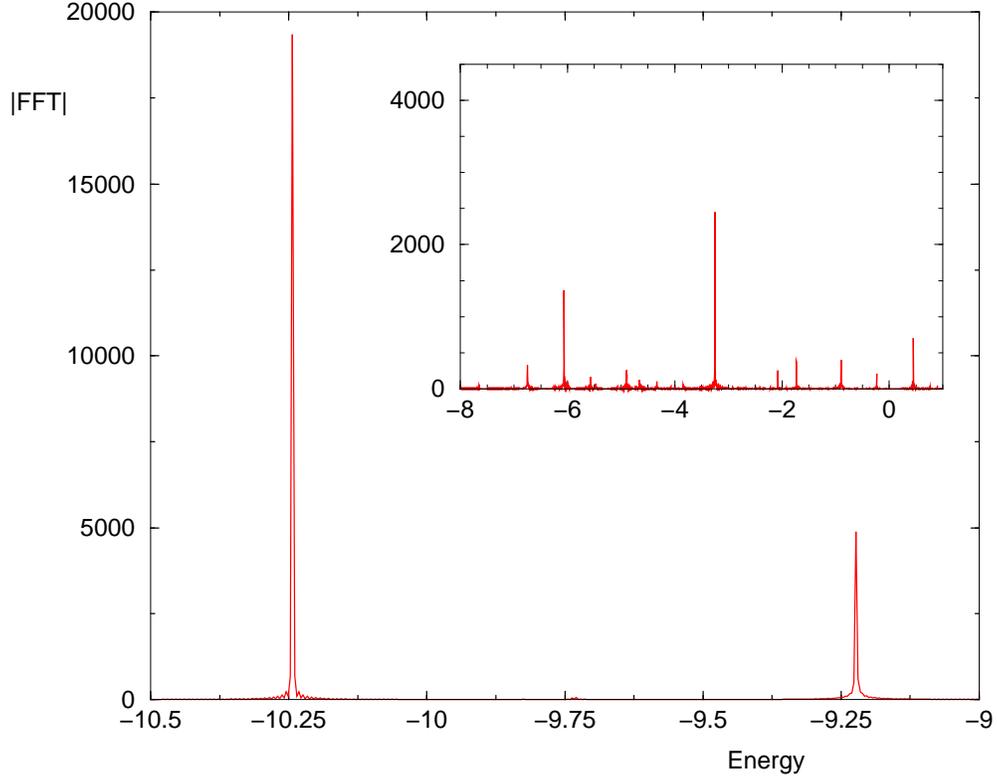}
\end{center}
\bigskip
\caption{Energy spectrum of the Hubbard model obtained from
the simulator. The lattice has $4\times 2$ sites (which requires 
16 qubits), with $t_x=1$, $t_y=1$
and ${\cal U}=4$ and the time steps used in the Trotter approximation (to prepare
the initial state and apply the evolution) are $\Delta t_1=\Delta t_2=0.05$.
%\mComment{Meaning of inset figure?}
}
\label{fig:14}
\end{figure}

In practice, to evaluate the discrete Fast Fourier Transform (DFFT) 
one uses discrete samples, therefore Eq. \ref {FFT} must be modified 
accordingly. In Fig.~\ref{fig:14} we see that instead of having 
$\delta$-functions (Dirac's functions), we have finite peaks in some 
range of energies, close to the eigenvalues of the Hamiltonian.
Accordingly, one cannot determine the eigenvalues with the same 
accuracy as other numerical calculations. However, there are some 
methods that give the results more accurately from the DFFT.

As a function of the frequency $\Omega_m$, the DFFT
($\tilde{F}(\Omega_m)$) is given by:
\begin{equation}
\tilde{F}(\Omega_m)=\Delta t \sum\limits_{j=0}^{N-1} F(t_j) 
e^{i \Omega_m t_j} \ ,
\end{equation}
where $t_j=j \Delta t$ are the different times at which the  function
$F$ is sampled (in the case of Section ~\ref{section3c4}, 
$F(t_j)=\langle \hat{U}(t_j) \rangle$),  $\Omega_m=\frac{2 \pi m}{N \Delta
t}$ are the possible frequencies to evaluate the FFT of $F(t)$ and $N$
is the number of samples. ($N$ must be an integer power of 2.)

Since we are interested in
$F(t)=\sum\limits_{n=0}^{\cal L}|\gamma_n|^2 \ e^{-i\lambda_nt}$ (Eq.
\ref{FFT0}),
\begin{equation}
\tilde{F}(\Omega_m)=\Delta t\sum\limits_{n=0}^{\cal L} 
|\gamma_n|^2 \sum\limits_{j=0}^{N-1}
e^{i[\Omega_m-\lambda_n]t_j} \ ,
\end{equation}
and then
\begin{equation}
\label{serie}
\tilde{F}(\Omega_m)=\Delta t\sum\limits_{n=0}^{\cal L} |\gamma_n|^2 
\ \frac{e^{i (\Omega_m -
\lambda_n)\Delta t N}-1}{e^{i (\Omega_m -\lambda_n)\Delta t}-1} \ .
\end{equation}
If $\Omega_m$ is close to one of the eigenvalues $\lambda_n$ and
the $\lambda_n$ are sufficiently far appart to be well resolved, we can
neglect all terms in the sum other than $n$.  If we take $\Omega_m$
and $\Omega_{m+1}=\Omega_m+\frac{2\pi}{N \Delta t}$, 
both close to $\lambda_n$ in such a way that
$|\tilde{F}(\Omega_m)|\gg |\tilde{F}(\Omega_{m+1})|\gg 0$, then from
Eq. \ref {serie} we find that
\begin{equation}
\label{coef}
\frac{\tilde{F}(\Omega_{m+1})}{\tilde{F}(\Omega_m)} \approx
\frac{e^{i (\Omega_m-\lambda_n)
\Delta t}-1}
{e^{i (\Omega_{m+1}-\lambda_n)\Delta t}-1} \ .
\end{equation}
After simple algebraic manipulations (and approximating 
$\ln(1+z) \approx z$ for $|z| \rightarrow 0$) we obtain the correction to
the energy $\lambda_n$:
\begin{equation}
\lambda_n=\Omega_m+\Delta\lambda_n
\end{equation}
with
\begin{equation}
\Delta\lambda_n \approx {\rm Re}\biggl[i\frac{\tilde{F}
(\Omega_{m+1})}{\tilde{F}(\Omega_m)}
\Bigl[\frac{e^{i2\pi/N}-1}{\Delta t}\Bigr]\biggr] 
\end{equation}

% BEGIN REFERENCES

\newpage
% FIGURES

\newpage

\newpage

\newpage

\newpage

\newpage

\newpage

\newpage

\newpage

\newpage

\newpage

\newpage

\newpage

\newpage

\newpage

\begin{thebibliography}{10}

%%%%%%%% section I
\bibitem{sfer}
G. Ortiz, J.E. Gubernatis, E. Knill, and R. Laflamme, Phys. Rev. A {\bf
64}, 22319 (2001).

\bibitem{somaroo:qc1999a}
S. Somaroo, C.-H. Tseng, T. F. Havel, R. Laflamme and D. G. Cory,
Phys. Rev. Lett. {\bf 82}, 5318 (1999).

\bibitem{batista01} 
C.D. Batista and G. Ortiz, Phys. Rev. Lett. {\bf 86}, 1082 (2001).  
C.D. Batista and G. Ortiz, cond-mat/xxxxx. 

\bibitem{terhal:qc1998a}
B.~M. Terhal and D.~P. DiVincenzo, Phys. Rev. A {\bf 61}, 2301 (2000).

\bibitem{qmcrev}
See for instance {\em Quantum Monte Carlo Methods in Condensed-Matter 
Physics,\/} edited  by M. Suzuki (World Scientific, Singapore, 1993);
W. M. C. Foulkes, L. Mitas,  R. J. Needs, and G. Rajagopal, Rev. Mod.
Phys. {\bf 73}, 33 (2001).

\bibitem{wiese}
We thank U.-J. Wiese and S. Chandrasekharan for pointing this out to us. 

\bibitem{farhi01}
E. Farhi, J. Goldstone, S. Gutmann, J. Lapan, A. Lundgren, and D. Preda,
Science {\bf 292}, 472 (2001).

%%%%%%%% section II
\bibitem{barenco:qc1995a}
A. Barenco {\it et~al.}, Phys. Rev. A {\bf 52},  3457  (1995).

\bibitem{divincenzo:qc1995a}
D. DiVincenzo, Phys. Rev. A {\bf 51},  1015  (1995).

\bibitem{price:qc1999a}
M. D. Price, S. S. Somaroo, A. E. Dunlop, T. F.
Havel, D. G. Cory, Phys. Rev. A {\bf 60}, 2777 (1999).

\bibitem{somaroo:qc1998a}
S. S. Somaroo, D. G. Cory, T. F. Havel, Phys. Lett. A {\bf 240}, 1 (1998).

\bibitem{bravyi:qc2000a}
S. Bravyi and A. Kitaev, quant-ph/0003137 (2000).

\bibitem{jordan}
P. Jordan and E. Wigner, Z. Phys. {\bf 47},  631  (1928).

\bibitem{fradkin}
E. Fradkin, Phys. Rev. Lett. {\bf 63},  322  (1989).

\bibitem{huerta}
L. Huerta and J. Zanelli, Phys. Rev. Lett. {\bf 71},  3622  (1993).

%%%%%%%% section III
\bibitem{blaizot}
J.-P. Blaizot and G. Ripka, {\em Quantum Theory of Finite Systems,\/}
(MIT Press, Cambridge, 1986).

\bibitem{peregilmo}
A. Perelomov, {\em Generalized Coherent States and their Applications} 
(Springer-Verlag, Berlin, 1986).

\bibitem{suzuki}
For a brief review, see M. Suzuki, in {\em Quantum Monte Carlo Methods 
in Condensed-Matter Physics,\/} edited by M. Suzuki (World Scientific,
Singapore, 1993), pg. 1.

\bibitem{paz}
C. Miquel, J. P. Paz, M. Saraceno, E. Knill, R. Laflamme and C. Negrevergne,
%``State Tomography and Spectroscopy as Quantum Computations'',
unpublished manuscript (2001).

\bibitem{kitaev}
A. Yu Kitaev, 
%``Quantum Measurement and the Abelian Stabilizer Problem'',
quant-ph/9511026 (1995).

\bibitem{cleve}
R. Cleve, A. Ekert, C. Macchiavello and M. Mosca,
Proc. R. Soc. Lond. A {\bf 454}, 339 (1998).

\bibitem{abrams}
D. S. Abrams and S. Lloyd, Phys. Rev. Lett. {\bf 83}, 5162 (1999).

\bibitem{negele:qc1988a}
J.~W. Negele and H. Orland, {\em Quantum Many-Particle Systems}
(Addison-Wesley, Redwood City, 1988).

\bibitem{bonca}
See for instance J. Bon$\breve{\rm c}$a and J. E. Gubernatis, Phys. Rev.
B {\bf 58}, 6992 (1998).

\bibitem{villars}
F. Villars, in {\em Proceedings of the International School of 
Physics ``Enrico Fermi'', Course XXIII: Nuclear Physics}, 
edited by V.F. Weisskopf (Academic Press, New York, 1963), pg. 1.


%%%%%%%% section IV

%\bibitem{shor:qc1995a}
%P.~W. Shor, SIAM J. Comput. {\bf 26},  1484  (1997).
%
%\bibitem{grover:qc1997a}
%L.~K. Grover, Phys. Rev. Lett. {\bf 79},  325  (1997).

%\bibitem{feynman:qc1982a}
%R.~P. Feynman, Int. J. Theor. Phys. {\bf 21},  467  (1982).

%\bibitem{abrams:qc1997a}
%D.~S. Abrams and S. Lloyd, Phys. Rev. Lett. {\bf 79},  2586  (1997).

%\bibitem{ours}
%C.~D. Batista and G. Ortiz, Phys. Rev. Lett. {\bf 86}, 1082 (2001).

%\bibitem{deutsch:qc1985a}
%D. Deutsch, Proc. R. Soc. Lond. A {\bf 400},  97  (1985).

%\bibitem{onsager}
%L. Onsager, Phys. Rev. {\bf 65},  117  (1944).

%\bibitem{bravyi:qc2000a}
%S. Bravyi and A. Kitaev, quant-ph/0003137 (unpublished).

%\bibitem{feynman:qc1965a}
%R.~P. Feynman and A.~R. Hibbs, {\em Quantum Mechanics and Path Integrals}
%  (McGraw-Hill, New York, 1965).

%\bibitem{Note0}
%The system is composed of $N_e$ particles moving in $d$ spatial dimensions
%  ($\hbar=m=e=1$), and a generic point in a flat Cartesian manifold of
%  dimension $D=d N_e$ is represented by ${\cal R} = ({\bf r}_1, \cdots, {\bf
%  r}_{N_e})$. $V({\cal R})$ is the potential energy operator and $p_i$ is
%  particle's $i$ canonical momentum.

%\bibitem{vonderlinden:qc1992a}
%W. von~der Linden, Phys. Rep. {\bf 220},  53  (1992).

%\bibitem{lloyd:qc1996a}
%S. Lloyd, Science {\bf 273},  1073  (1996);
%D. A. Meyer, J. Stat. Phys. {\bf 85}, 551 (1996).
%S. Wiesner, quant-ph/9603028;
%B. M. Boghosian and W. Taylor IV, Physica D {\bf 120}, 30 (1998).
%C. Zalka, Proc. R. Soc. Lond. A {\bf 454}, 313 (1998). 

%\bibitem{knill:qc1999c}
%E. Knill and R. Laflamme, quant-ph/9909094 (unpublished).

%\bibitem{Note2}
%From the theory of computation point of view it is necessary to make additional
%  assumptions on how the functions may be prescribed. In particular the
%  functions themselves must be classically computable in a suitable sense. This
%  problem is avoided by permitting only a finite set of quantum gates instead
%  of continuously controllable Hamiltonians.

%\bibitem{peres:qc1998a}
%A. Peres, {\em Quantum Theory: Concepts and Methods} (Kluwer Academic
%  Publishers, Dordrecht, The Netherlands, 1998).

%\bibitem{bernstein:qc1997a}
%E. Bernstein and U. Vazirani, SIAM J. Comput. {\bf 26},  1411  (1997).

%\bibitem{cleve:qc1999b}
%R. Cleve, quant-ph/9906111 (unpublished).

%\bibitem{aharonov:qc1998b}
%D. Aharonov, quant-ph/9812037 (unpublished).

%\bibitem{guerrero:qc1999a}
%M. Guerrero, G. Ortiz, and J.~E. Gubernatis, Phys. Rev. B {\bf 59},  1706
%  (1999).

%\bibitem{kitaev:qc1995a}
%A.~Y. Kitaev, quant-ph/9511026 (unpublished).

%\bibitem{terhal:qc1998a}
%B.~M. Terhal and D.~P. DiVincenzo, Phys. Rev. A {\bf 61},  2301  (2000).

%\bibitem{knill:qc1998c}
%E. Knill and R. Laflamme, Phys. Rev. Lett. {\bf 81},  5672  (1998).

%\bibitem{shor:qc1995b}
%P.~W. Shor, Phys. Rev. A {\bf 52},  2493  (1995).

%\bibitem{steane:qc1995a}
%A. Steane, Proc. R. Soc. Lond. A {\bf 452},  2551  (1996).

%\bibitem{shor:qc1996a}
%P.~W. Shor,  in {\em Proceedings of the Symposium on the Foundations of
%  Computer Science}, (IEEE press, Los Alamitos, California, 1996),
%  pp.\ 56--65.

%\bibitem{aharonov:qc1996a}
%D. Aharonov and M. Ben-Or,  in {\em Proceedings of the 29th Annual ACM
%  Symposium on the Theory of Computation (STOC)} (ACM Press, New York, New
%  York, 1996), pp.\ 176--188.

%\bibitem{kitaev:qc1996a}
%A.~Y. Kitaev,  in {\em Quantum Communication, Computing and Measurement},
%  edited by O. Hirota, A.~S. Holevo, and C.~M. Caves (Plenum Press, New York,
%  1997).

%\bibitem{knill:qc1998a}
%E. Knill, R. Laflamme, and W.~H. Zurek, Science {\bf 279},  342  (1998).

%\bibitem{preskill:qc1998a}
%J. Preskill, Proc. R. Soc. Lond. A {\bf 454},  385  (1998).

%\bibitem{gellmann}
%M. Gell-Mann and F. Low, Phys. Rev. {\bf 84},  350  (1951).

\end{thebibliography}
\end{document}